\documentclass[twocolumn, twocolappendix]{aastex63}

\usepackage{bm} % bold math
\usepackage{chemformula} % for chemical reactions and molecules naming
\let\ce\ch % for chemical reactions and molecules naming
\usepackage{xspace} % for correct shortcuts spacing from other words and punctuation marks
\newcommand{\kms}{\,km\,s$^{-1}$\xspace} % kilometres per second
\newcommand{\mm}{\,mm\xspace} % millimeters
\newcommand{\K}{\,K\xspace} % Kelvin
\newcommand{\mK}{\,mK\xspace} % millikelvin
\newcommand{\kK}{\,kK\xspace} % kilokelvin
\newcommand{\GHz}{\,GHz\xspace} % Gigahertz
\newcommand{\kHz}{\,kHz\xspace} % Kilohertz
\newcommand{\D}{\,D\xspace} % Debyes
\usepackage{booktabs} % proof for aligning in table
\usepackage{array} % proof for aligning in table
\usepackage{upgreek} % Greek leters not in italics

\received{2023 December 24}
\revised{2024 March 21}
\accepted{2024 April 3}
\shorttitle{First detection of \ce{H2CNCN} in the ISM}
\shortauthors{San Andr{\'e}s et al.}
\begin{document}

\title{First detection in space of the high-energy isomer of cyanomethanimine: \ce{H2CNCN}}

\correspondingauthor{David San Andr{\'e}s}
\email{david.sanandres@cab.inta-csic.es}

\author[0000-0001-7535-4397]{David San Andr{\'e}s}
\affiliation{Centro de Astrobiolog{\'i}a (CAB), INTA-CSIC, Carretera de Ajalvir km 4, Torrej{\'o}n de Ardoz, 28850, Madrid, Spain}

\author[0000-0002-2887-5859]{V{\'i}ctor M. Rivilla}
\affiliation{Centro de Astrobiolog{\'i}a (CAB), INTA-CSIC, Carretera de Ajalvir km 4, Torrej{\'o}n de Ardoz, 28850, Madrid, Spain}

\author[0000-0001-8064-6394]{Laura Colzi}
\affiliation{Centro de Astrobiolog{\'i}a (CAB), INTA-CSIC, Carretera de Ajalvir km 4, Torrej{\'o}n de Ardoz, 28850, Madrid, Spain}

\author[0000-0003-4493-8714]{Izaskun Jim{\'e}nez-Serra}
\affiliation{Centro de Astrobiolog{\'i}a (CAB), INTA-CSIC, Carretera de Ajalvir km 4, Torrej{\'o}n de Ardoz, 28850, Madrid, Spain}

\author[0000-0003-4561-3508]{Jes{\'u}s Mart{\'i}n-Pintado}
\affiliation{Centro de Astrobiolog{\'i}a (CAB), INTA-CSIC, Carretera de Ajalvir km 4, Torrej{\'o}n de Ardoz, 28850, Madrid, Spain}

\author[0000-0002-6389-7172]{Andr{\'e}s Meg{\'i}as}
\affiliation{Centro de Astrobiolog{\'i}a (CAB), INTA-CSIC, Carretera de Ajalvir km 4, Torrej{\'o}n de Ardoz, 28850, Madrid, Spain}

\author[0000-0001-6049-9366]{{\'A}lvaro L{\'o}pez-Gallifa}
\affiliation{Centro de Astrobiolog{\'i}a (CAB), INTA-CSIC, Carretera de Ajalvir km 4, Torrej{\'o}n de Ardoz, 28850, Madrid, Spain}

\author[0000-0001-5191-2075]{Antonio Mart{\'i}nez-Henares}
\affiliation{Centro de Astrobiolog{\'i}a (CAB), INTA-CSIC, Carretera de Ajalvir km 4, Torrej{\'o}n de Ardoz, 28850, Madrid, Spain}

\author[0000-0002-7387-9787]{Sarah Massalkhi}
\affiliation{Centro de Astrobiolog{\'i}a (CAB), INTA-CSIC, Carretera de Ajalvir km 4, Torrej{\'o}n de Ardoz, 28850, Madrid, Spain}

\author[0000-0003-3721-374X]{Shaoshan Zeng}
\affiliation{Star and Planet Formation Laboratory, Cluster for Pioneering Research, RIKEN, 2–1 Hirosawa, Wako, Saitama, 351–0198, Japan}

\author[0000-0001-9629-0257]{Miguel Sanz-Novo}
\affiliation{Centro de Astrobiolog{\'i}a (CAB), INTA-CSIC, Carretera de Ajalvir km 4, Torrej{\'o}n de Ardoz, 28850, Madrid, Spain}

\author[0000-0002-4782-5259]{Bel{\'e}n Tercero}
\affiliation{Observatorio Astron{\'o}mico Nacional (OAN-IGN), Calle Alfonso XII, 3, 28014 Madrid, Spain}
\affiliation{Observatorio de Yebes (OY-IGN), Cerro de la Palera SN, Yebes, Guadalajara, Spain}

\author[0000-0002-5902-5005]{Pablo de Vicente}
\affiliation{Observatorio de Yebes (OY-IGN), Cerro de la Palera SN, Yebes, Guadalajara, Spain}

\author[0000-0001-9281-2919]{Sergio Mart{\'i}n}
\affiliation{European Southern Observatory, Alonso de C{\'o}rdova, 3107, Vitacura, Santiago 763-0355, Chile}
\affiliation{Joint ALMA Observatory, Alonso de C{\'o}rdova, 3107, Vitacura, Santiago 763-0355, Chile}

\author[0009-0009-5346-7329]{Miguel Angel Requena Torres}
\affiliation{Department of Physics, Astronomy, and Geosciences, Towson University, Towson, MD 21252, USA}

\author[0000-0001-8803-8684]{Germ{\'a}n Molpeceres}
\affiliation{Department of Astronomy, Graduate School of Science, The University of Tokyo, Tokyo 113 0033, Japan}

\author[0000-0001-6484-9546]{Juan Garc{\'i}a de la Concepci{\'o}n}
\affiliation{Departamento de Qu{\'i}mica Org{\'a}nica e Inorg{\'a}nica, Facultad de Ciencias, Universidad de Extremadura, E-06006 Badajoz, Spain}
\affiliation{IACYS-Unidad de Qu{\'i}mica Verde y Desarrollo Sostenible, Facultad de Ciencias, Universidad de Extremadura, E-06006 Badajoz, Spain}

%% Note that the \and command from previous versions of AASTeX is now
%% depreciated in this version as it is no longer necessary. AASTeX 
%% automatically takes care of all commas and "and"s between authors names.

%% AASTeX 6.3 has the new \collaboration and \nocollaboration commands to
%% provide the collaboration status of a group of authors. These commands 
%% can be used either before or after the list of corresponding authors. The
%% argument for \collaboration is the collaboration identifier. Authors are
%% encouraged to surround collaboration identifiers with ()s. The 
%% \nocollaboration command takes no argument and exists to indicate that
%% the nearby authors are not part of surrounding collaborations.

%% Mark off the abstract in the ``abstract'' environment. 
\vspace{1cm}
\begin{abstract}

We report the first detection in the interstellar medium of $N$-cyanomethanimine (\ce{H2CNCN}), the stable dimer of \ce{HCN} of highest energy, and the most complex organic molecule identified in space containing the prebiotically relevant \ce{NCN} backbone. We have identified a plethora of $a$-type rotational transitions with $3 \leq J_\text{up} \leq 11$ and $K_\text{a} \leq 2$ that belong to this species towards the Galactic Center G+0.693-0.027 molecular cloud, the only interstellar source showing the three cyanomethanimine isomers (including the $Z$- and $E$- isomers of $C$-cyanomethanimine, \ce{HNCHCN}). We have derived a total column density for \ce{H2CNCN} of (2.9$\, \pm \,$0.1)$\times10^{12} \; \text{cm}^{-2}$, which translates into a total molecular abundance with respect to \ce{H2} of (2.1$\, \pm \,$0.3)$\times10^{-11}$. We have also revisited the previous detection of $E$- and $Z$-\ce{HNCHCN}, and found a total $C/N$-cyanomethanimine abundance ratio of 31.8$\, \pm \,$1.8 and a $Z/E$-\ce{HNCHCN} ratio of 4.5$\, \pm \,$0.2. While the latter can be explained on the basis of thermodynamic equilibrium, chemical kinetics are more likely responsible for the observed $C/N$-cyanomethanimine abundance ratio, where the gas-phase reaction between methanimine (\ce{CH2NH}) and the cyanogen radical (\ce{CN}) arises as the primary formation route.

%This example manuscript is intended to serve as a tutorial and template for
%authors to use when writing their own AAS Journal articles. The manuscript
%includes a history of \aastex\ and documents the new features in the
%previous versions as well as the new features in version 6.3. This
%manuscript includes many figure and table examples to illustrate these new
%features.  Information on features not explicitly mentioned in the article
%can be viewed in the manuscript comments or more extensive online
%documentation. Authors are welcome replace the text, tables, figures, and
%bibliography with their own and submit the resulting manuscript to the AAS
%Journals peer review system.  The first lesson in the tutorial is to remind
%authors that the AAS Journals, the Astrophysical Journal (ApJ), the
%Astrophysical Journal Letters (ApJL), and Astronomical Journal (AJ), all
%have a 250 word limit for the abstract.  If you exceed this length the
%Editorial office will ask you to shorten it. This abstract has 180 words.

\end{abstract}

%% Keywords should appear after the \end{abstract} command. 
%% See the online documentation for the full list of available subject
%% keywords and the rules for their use.

%% From the front matter, we move on to the body of the paper.
%% Sections are demarcated by \section and \subsection, respectively.
%% Observe the use of the LaTeX \label
%% command after the \subsection to give a symbolic KEY to the
%% subsection for cross-referencing in a \ref command.
%% You can use LaTeX's \ref and \label commands to keep track of
%% cross-references to sections, equations, tables, and figures.
%% That way, if you change the order of any elements, LaTeX will
%% automatically renumber them.
%%
%% We recommend that authors also use the natbib \citep
%% and \citet commands to identify citations.  The citations are
%% tied to the reference list via symbolic KEYs. The KEY corresponds
%% to the KEY in the \bibitem in the reference list below. 

%%%%%%%%%%%%%%%%%%%%%%%%%%%%%%%%%%%%%%%%%%%%%%%%%%%%%%%%%%
%%%%%%%%%%%%%%%%%%%%%% INTRODUCTION %%%%%%%%%%%%%%%%%%%%%%
%%%%%%%%%%%%%%%%%%%%%%%%%%%%%%%%%%%%%%%%%%%%%%%%%%%%%%%%%%

\section{Introduction} \label{sec:intro}

How life originated on Earth is one of the key open questions still lingering in the astrobiological community. One of the most accepted hypothesis is the so-called RNA-world \citep{Gilbert1986}, which suggests that this macromolecule may have performed both the metabolic and genetic functions that in present-day living organisms are carried out by proteins and DNA, respectively. Although the chemical processes forming the first RNA molecules remain unknown, numerous prebiotic chemistry experiments have shown that RNA building blocks, the ribonucleotides, can be synthesised from much simpler molecules (\citealt{Powner2009, Patel2015, Becker2016, Becker2019}), with nitriles (organic species with the \ce{-C+N} functional group) playing a dominant role (e.g., \citealt{Balucani2009, Menor-Salvan2020}). The clearest example is found on their simplest representative, hydrogen cyanide (\ce{HCN}), which has been proven to be an essential ingredient in initiating many prebiotic synthesis mechanisms (e.g., \citealt{Oro&Kimball1961, Ferris&Hagan1984, Santalucia2022, Sandstrom&Rahm2023}). 
Among them, \ce{HCN} oligomerization reactions are specially relevant, as these are evidenced to be the major routes forming the two purine nucleobases assembling both RNA and DNA chains: adenine (\ce{H5C5N5}; \citealt{Chakrabarti&Chakrabarti2000, Jung&Choe2013}), and guanine (\ce{H5C5N5O}; \citealt{Sanchez1968,Choe2018}). Since \ce{HCN} is ubiquitous in the interstellar medium (ISM), these processes might similarly occur through interstellar chemistry, making the study of the molecular complexity in the ISM an essential task to gain a deeper insight into this ``vital'' question.

A crucial step triggering the \ce{HCN} oligomers cascade concerns the initial formation of \ce{HCN} dimers (\ce{H2C2N2}).
The formation of such species from two separate \ce{HCN} molecules did not appear to be efficient under the typical temperatures of the ISM ($\sim$10$-$100\K) due to the large activation energy barrier involved ($\sim$36000\K; \citealt{Smith2001, Yim&Choe2012}). However, the successful detections in the ISM of the two most stable \ce{HCN} dimers, the $E$- and $Z$- isomers of $C$-cyanomethanimine (\ce{HNCHCN}; \citealt{Zaleski2013} and \citealt{Rivilla2019}, respectively), triggered the study of alternative chemical pathways (e.g., \citealt{Vazart2015, Shivani2017,Shingledecker2020,GarciadelaConcepcion2021}).

In this regard, the exhaustive characterisation of new \ce{HCN} dimers in the ISM would substantially contribute to a better understanding on how these compounds could be formed in the interstellar environment.
From the entire catalog of possible \ce{HCN} dimers, only a single additional isomer has shown superior stability compared to two isolated \ce{HCN} molecules \citep{Evans1991}. That is $N$-cyanomethanimine (\ce{H2CNCN}), a higher-energy member within this family (3570$\, \pm \,$130\K above the $Z$-isomer, which is the global minimum; \citealt{Puzzarini2015}), whose identification in the ISM has so far remained elusive. This species is of prime astrobiological interest since, besides being a \ce{HCN} dimer, it contains the \ce{NCN} backbone, a fundamental structure of purine nucleobases which is only present in a handful of simpler interstellar molecules: cyanamide (\ce{NH2CN}; \citealt{Turner1975}), carbodiimide (\ce{HNCNH}; \citealt{McGuire2012}), isocyanogen (\ce{CNCN}; \citealt{Agundez2018}) and the cyanomidyl radical (\ce{HNCN}; \citealt{Rivilla2021b}). Furthermore, recent experiments conducted by \citet{Vasconcelos2020} have proposed an alternative route for adenine formation within interstellar dust grain ices that relies on \ce{H2CNCN} as the primary precursor, so its detection in the ISM would increase the chances of adenine being formed under the extreme interstellar conditions.

In this work, we present the first detection of several rotational lines of $N$-cyanomethanimine in the ISM towards the Galactic Center G+0.693-0.027 molecular cloud (hereafter G+0.693), the same source which also host the unique detection of both $C$-cyanomethanimine isomers. This cloud, which belongs to the Sgr B2 complex, emerges as one of the most promising candidates for the search of complex organic molecules (COMs; \citealt{Herbst2020}). Its chemistry is believed to be affected by low-velocity shocks likely driven by large-scale cloud-cloud collisions, sputtering molecules that are formed in the surface of dust grains into the gas-phase \citep{Martin2008, Zeng2020}. These particular conditions have endowed it with an unprecedented chemical richness, portrayed in the more than 130 molecular species that have been already identified towards it, many of them of high prebiotic interest (e.g., \citealt{Zeng2019, Zeng2021, Zeng2023, Rivilla2019, Rivilla2022c, Rivilla2023, Jimenez-Serra2020, Jimenez-Serra2022, Rodriguez-Almeida2021a, Rodriguez-Almeida2021b}). 

This paper is organised as follows. In Sect.~\ref{sec:observations} we present the observational data,
while in Sect.~\ref{sec:analysis_and_results} we report the detection of $N$-cyanomethanimine towards G+0.693, and revisit the analysis of the two isomers of $C$-cyanomethanimine. In Sect.~\ref{sec:discussion} we discuss the origin of their observed relative isomeric ratios by means of their main formation and destruction chemical routes. Finally, in Sect.~\ref{sec:summary_and_conclusions} we outline our conclusions.

%%%%%%%%%%%%%%%%%%%%%%%%%%%%%%%%%%%%%%%%%%%%%%%%%%%%%%%%%%
%%%%%%%%%%%%%%%%%%%%%% OBSERVATIONS %%%%%%%%%%%%%%%%%%%%%%
%%%%%%%%%%%%%%%%%%%%%%%%%%%%%%%%%%%%%%%%%%%%%%%%%%%%%%%%%%

\section{OBSERVATIONS} \label{sec:observations}

\begin{deluxetable*}{cccccc}[ht!]
\tablenum{1}
\tablecaption{Derived physical parameters of the best LTE fit for the three cyanomethanimine species targeted in this work.\label{tab:fitting_parameters}}
\tablewidth{0pt}
\tablehead{
\colhead{Molecule} & \colhead{$N$} & \colhead{$T_\text{ex}$} & \colhead{$\text{FWHM}$} & \colhead{$v_\text{LSR}$} & \colhead{$N/N_{\text{\ce{H2}}}$$^{(a)}$} \\
\colhead{} & \colhead{$(\times 10^{12} \, \text{cm}^{-2})$} & \colhead{$\text{K}$} & \colhead{$(\text{km} \, \text{s}^{-1})$} & \colhead{$(\text{km} \, \text{s}^{-1})$} &
\colhead{$(\times 10^{-11})$}
}
\startdata
\ce{H2CNCN} & $2.9 \pm 0.1$ & $7.9 \pm 0.3$ & $21.9 \pm 1.4$ & $68.1 \pm 0.5$ & $2.1 \pm 0.3$ \\
$Z$-\ce{HNCHCN} & $75 \pm 3$ & $14.1 \pm 0.8$ & $22.3 \pm 0.8$ & $67.0 \pm 0.5$ & $55 \pm 9$ \\
$E$-\ce{HNCHCN} & $16.8 \pm 0.3$ & $15.5 \pm 0.5$ & $22.1 \pm 0.6$ & $67.8 \pm 0.2$ & $12 \pm 2$ \\
\enddata
\tablecomments{ The FWHM estimates were firstly determined by only fitting the most unblended lines of these species, while the remaining three parameters were obtained through the complete LTE modelling in which these FWHM values were fixed (see Sects.~\ref{subsec:detection_of_H2CNCN} and \ref{subsec:analysis_of_Z_E-HNCHCN}). Their uncertainties correspond to $1\sigma$ standard deviations. \\
$^{(a)}$ Fractional abundances with respect to \ce{H2}. To compute them, we used $N_{\text{\ce{H2}}} = 1.35 \times10^{23} \, \text{cm}^{-2}$ as derived by \citet{Martin2008}, while assuming an uncertainty of 15$\%$ of its value.}
\end{deluxetable*}

We have analysed data from the recently enhanced high-sensitivity spectral survey of G+0.693 \citep{Rivilla2023, Sanz-Novo2023}. New ultradeep observational runs were performed using the Yebes 40 m (Guadalajara, Spain; project 21A014, PI: Rivilla) and IRAM 30 m (Granada, Spain; project 123-22, PI: Jim{\'e}nez-Serra) radio telescopes, which have remarkably reduced the rms of the spectra in comparison with previous works (e.g., \citealt{Rivilla2020a, Rivilla2021a, Rivilla2021b, Rivilla2022c, Zeng2020, Rodriguez-Almeida2021a, Colzi2022}). The observations were centered at $\alpha_\text{J2000} = 17^\text{h}47^\text{m}22^\text{s}$, $\delta_\text{J2000} = -28^\text{o}21'27''$ and were conducted using position switching mode, being the off position located at an offset of $\Delta\alpha = -885''$, $\Delta\delta = 290''$. This particular reference position was selected because it does not show significant emission from abundant molecules such as \ce{CS} and \ce{HC3N}.
The line intensity of the spectra was directly measured in antenna temperature ($T_\text{A}^*$) units, as the molecular emission towards G+0.693 is extended over the beam (e.g., \citealt{Brunken2010, Jones2012, Li2020, Zheng2024}). 

For the Yebes 40 m observations, the Nanocosmos Q-band (7\mm) HEMT receiver was used to cover the whole Q-band frequency range (31.07$-$50.42\GHz), providing a raw frequency resolution of $\sim$38\kHz \citep{Tercero2021}. On the other hand, the new IRAM 30 m observations set was gathered by combining the broadband heterodyne Eight MIxer Receiver (EMIR) and the Fast Fourier Transform Spectrometer FTS200, providing a raw channel width of $\sim$195\kHz along the three spectral ranges covered: 83.20$-$115.41, 132.28$-$140.39 and 142.00$-$173.81\GHz. In both cases, the spectra were ultimately smoothed to achieve a final resolution of $\sim$256\kHz (1.5$-$2.5\kms) for the Yebes 40 m data and of $\sim$615\kHz (1.1$-$2.2\kms) for the IRAM 30 m observations, which is more than enough to resolve G+0.693 molecular line profiles that present typical line widths of $\sim$15$-$25\kms. The rms of the spectra ranges from 0.25$-$0.9\mK across the frequency range observed with the Yebes 40 m radio telescope, and between 0.5$-$2.5\mK and 1.0$-$1.6\mK for the IRAM 30 m datasets at 3\mm and 2\mm, respectively. The half power beam width (HPBW) of the Yebes 40 m telescope varies from $\sim$55$''$ at 31\GHz down to $\sim$35$''$ at 50\GHz, while the HPBW of the IRAM 30 m radio telescope ranges from $\sim$29$''$ down to $\sim$14$''$ along the frequency range observed with it. We refer to \citet{Rivilla2023} and \citet{Sanz-Novo2023} for further details on both the observations and the data reduction process. Data belonging to the spectral ranges unrelated to this new set of observations come from our previous IRAM 30 m survey \citep{Rivilla2021a, Rivilla2021b, Rivilla2022c}.

%%%%%%%%%%%%%%%%%%%%%%%%%%%%%%%%%%%%%%%%%%%%%%%%%%%%%%%%%%
%%%%%%%%%%%%%%%%%% ANALYSIS AND RESULTS %%%%%%%%%%%%%%%%%%
%%%%%%%%%%%%%%%%%%%%%%%%%%%%%%%%%%%%%%%%%%%%%%%%%%%%%%%%%%

\section{ANALYSIS AND RESULTS} \label{sec:analysis_and_results}

The identification of the three \ce{C2H2N2} isomers molecular lines and their fitting has been performed using the version from 2023 November 15 of the Spectral Line Identification and Modelling (SLIM) tool within the \textsc{MADCUBA} package \citep{Martin2019}. For each of these species, we incorporated the spectroscopic data available within the Cologne Database for Molecular Spectroscopy catalog (CDMS, \citealt{Endres2016}). We just considered the rotational transitions without taking into account the complex hyperfine structure arising from the presence of two \ce{^{14}N} nuclei, since it is not resolved at the observed spectral resolution (e.g., the maximum separation between the two \ce{H2CNCN} hyperfine components that reproduce the bulk of the emission of its most intense unblended transition at 41.81\GHz shown in Fig.~\ref{fig:H2CNCN_detected_lines} is $\sim$6.2\kHz or, equivalently, $\sim$0.04\kms; a difference that further collapses for those transitions at higher frequencies). Nonetheless, we did also carry out the analysis of the three isomers including the entries with hyperfine structure lines, not observing substantial modification in the results obtained.

The modelling of the line profiles of the \ce{C2H2N2} isomers has been conducted assuming local thermodynamic equilibrium (LTE) conditions, since collisional coefficients for these species have not yet been calculated. The SLIM code operates under this assumption, and generates a synthetic spectra to be compared with the observed one. To derive the physical parameters describing the molecular emission, we used the automatic fitting routine SLIM-AUTOFIT, which provides the best non-linear least-squares LTE fit to the data using the Levenberg-Marquardt algorithm. The free parameters fitting the spectra are the total column density of the molecule ($N$), the excitation temperature ($T_\text{ex}$), the local standard of rest velocity ($v_\text{LSR}$), and the full width at half maximum ($\text{FWHM}$). 

We have performed the LTE line fitting of the species of interest in this work by considering the already modelled emission of the more than 130 molecular species previously identified towards G+0.693 \citep{Requena-Torres2006, Requena-Torres2008, Rivilla2018, Rivilla2019, Rivilla2020b, Rivilla2021a, Rivilla2021b, Rivilla2022a, Rivilla2022b, Rivilla2022c, Rivilla2023, Zeng2018, Zeng2020, Zeng2021, Zeng2023, Bizzocchi2020, Jimenez-Serra2020, Jimenez-Serra2022, Rodriguez-Almeida2021a, Rodriguez-Almeida2021b, Colzi2022, Alberton2023, Fatima2023, Massalkhi2023, SanAndres2023, Sanz-Novo2023}. In the forthcoming sections, we provide the details on the analysis for each cyanomethanimine species, whose results are summarised in Table~\ref{tab:fitting_parameters}.

\subsection{Detection of \ce{H2CNCN}} \label{subsec:detection_of_H2CNCN}

\begin{figure*}[ht!]
    \centering
    \includegraphics[width=\textwidth]{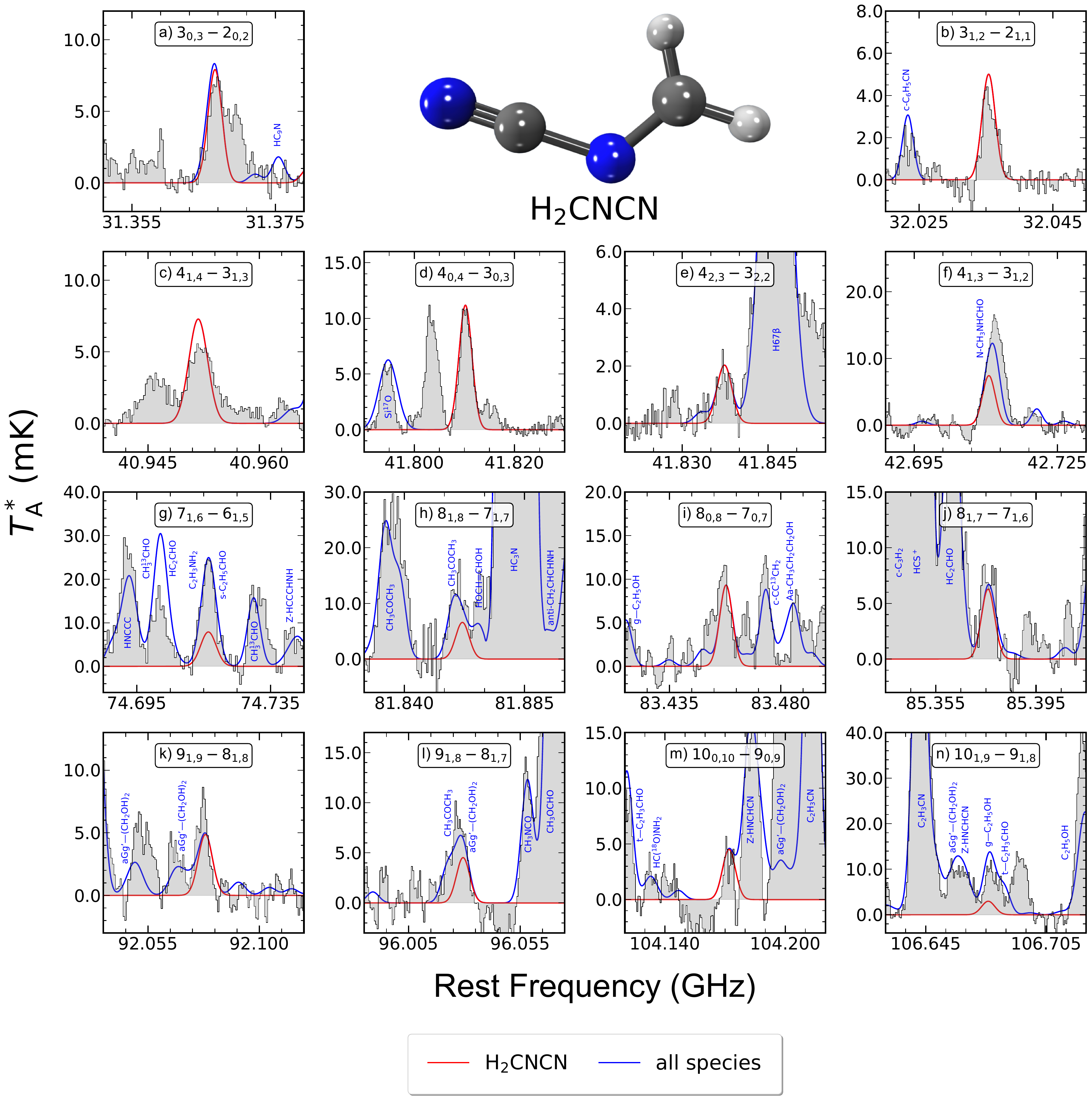}
    \caption{Unblended or partially blended \ce{H2CNCN} transitions detected towards G+0.693. The black histogram and the grey-shaded areas indicate the observed spectrum, while the red and blue solid lines represent the best LTE fit obtained for \ce{H2CNCN}, and the emission of all the species already identified in this cloud (which are indicated by the blue labels), respectively. Panel labels indicate the \ce{H2CNCN} rotational transition being plotted using the common notation for asymmetric tops: $J_{K_\text{a}K_\text{c}}$, where $J$ refers to the total angular momentum of the molecule while $K_\text{a,c}$ point to its projections along the $a$ and $c$ principal axes. Transitions shown here are those which have been used to perform the LTE line fitting of this molecule, and their spectroscopic information is given in Table~\ref{tab:H2CNCN_spectroscopy}. The molecular structure of \ce{H2CNCN} is depicted in the upper middle part (carbon atoms in dark-grey, nitrogen atoms in blue and hydrogen atoms in white colors).} 
    \label{fig:H2CNCN_detected_lines} 
\end{figure*}

\begin{deluxetable*}{DCDDcccc}[ht!]
\tablenum{2}
\tabletypesize{\footnotesize}
\tablecaption{List of \ce{H2CNCN} rotational transitions detected towards G+0.693 and selected for this molecule LTE fit.\label{tab:H2CNCN_spectroscopy}}
\setlength{\tabcolsep}{3.4pt}
\tablehead{
\multicolumn2c{Frequency} & \colhead{Transition$^{(a)}$} & \multicolumn2c{$\log I$} & \multicolumn2c{$E_\text{up}$} & \colhead{rms} & \colhead{$\int{T^*_\text{A}\text{d}v}$} & \colhead{S/N$^{(b)}$} & \colhead{Blending$^{(c)}$} \\
\multicolumn2c{($\text{GHz}$)} & \colhead{($J_{K_\text{a},K_\text{c}}$$'-$$J_{K_\text{a},K_\text{c}}$$''$)} & \multicolumn2c{($\text{nm}^2 \, \text{MHz}$)} & \multicolumn2c{($\text{K}$)} & \colhead{($\text{mK}$)} & \colhead{($\text{mK} \, \text{km} \, \text{s}^{-1}$)} & \colhead{} & \colhead{}
}
\decimals
\startdata
31.3664834(10) & 3_{0,3}-2_{0,2} & -5.469 & 3.0 & 0.5 & 179(15) & 53(5) & Blended with U \\
32.0353082(10) & 3_{1,2}-2_{1,1} & -5.505 & 5.9 & 0.5 & 113(13) & 34(4) & Unblended \\
40.9516386(13) & 4_{1,4}-3_{1,3} & -5.183 & 7.7 & 0.5 & 165(15) & 56(5) & Unblended \\
41.8100713(12) & 4_{0,4}-3_{0,3} & -5.133 & 5.0 & 0.5 & 253(19) & 87(7) & Unblended \\
41.8372354(12) & 4_{2,3}-3_{2,2} & -5.444 & 16.2 & 0.5 & 46(11) & 16(4) & Unblended \\
42.7104885(13) & 4_{1,3}-3_{1,2} & -5.147 & 7.9 & 0.5 & 168(15) & 58(6) & Blended: $N$-\ce{CH3NHCHO} and U \\
74.7161692(19) & 7_{1,6}-6_{1,5} & -4.468 & 17.1 & 2.4 & 180(60) & 10(3) & Blended: $s$-\ce{C2H5CHO} and \ce{C2H3NH2} \\
81.8614891(21) & 8_{1,8}-7_{1,7} & -4.344 & 20.5 & 2.4 & 151(60) & 9(3) & Blended: \ce{CH3COCH3} \\
83.4576038(20) & 8_{0,8}-7_{0,7} & -4.317 & 18.0 & 1.6 & 213(40) & 19(4) & Blended with U \\
85.3755815(21) & 8_{1,7}-7_{1,6} & -4.309 & 21.2 & 1.6 & 144(40) & 13(3) & Unblended$^\star$ \\
92.0779012(21) & 9_{1,9}-8_{1,8} & -4.204 & 24.9 & 0.9 & 112(22) & 19(4) & Blended with U \\
96.0290895(21) & 9_{1,8}-8_{1,7} & -4.169 & 25.8 & 1.5 & 104(35) & 11(4) & Blended: $aGg'$-\ce{(CH2OH)2} and \ce{CH3COCH3} \\
104.171562(10) & 10_{0,10}-9_{0,9} & -4.057 & 27.5 & 1.8 & 104(42) & 9(4) & Slightly blended: $Z$-\ce{HNCHCN} \\
106.675828(10) & 10_{1,9}-9_{1,8} & -4.045 & 31.0 & 1.8 & 69(41) & 6(3) & Blended: $g$-\ce{C2H5OH} \\
\enddata
\tablecomments{ For each transition,  we provide its rest frequency in units of $\text{GHz}$, its associated quantum numbers, the base 10 logarithm of its integrated intensity at a fixed temperature of 300\K in units of $\text{nm}^2 \, \text{MHz}$ ($\log I$), the energy in $\text{K}$ of the upper level involved in the transition ($E_\text{up}$), the noise measured in $\text{mK}$ within line-free spectral ranges close to it (rms), the integrated intensity over the line width as derived from the fit in $\text{mK}\,\text{km}\,\text{s}^{-1}$ ($\int{T^*_\text{A}\text{d}v}$), and the detection level (in terms of the signal-to-noise or S/N ratio). The numbers in brackets represent the uncertainty associated to the last digits. The last column accounts for the possible contamination of other molecular species. \\
$^{(a)}$ Rotational transitions are labelled following the common notation for asymmetric tops: $J_{K_\text{a}, K_\text{c}}$, where $J$ refers to the total angular momentum of the molecule while $K_\text{a,c}$ point to its projections along the $a$ and $c$ principal axes. \\
$^{(b)}$ The S/N is calculated from the integrated intensity over the line width ($\int{T^*_\text{A}\text{d}v}$) and noise level $\sigma = \text{rms}\times\sqrt{\updelta v\times\text{FWHM}}$, where $\updelta v$ is the spectral resolution of the spectra in velocity units and the $\text{FWHM}$ is estimated from the LTE line fitting. \\
$^{(c)}$ The term ``unblended'' alludes to those transitions which exhibit no contamination of other molecular species, while a star symbol ($\star$) is added when scarce line blending accounting for less than 5\% of the targeted line total integrated intensity is present. ``U'' refers to blending with an unknown (not yet identified) species.}
\end{deluxetable*}

$N$-cyanomethanimine (\ce{H2CNCN}) is a planar asymmetric-top molecule with a large total dipole moment of 4.84\D \citep{Bak1978, Bak&Svanholt1980}. To perform its search towards G+0.693, we used the spectroscopic entry 054514 (February 2014) from the CDMS catalog, which incorporates the rotational spectroscopy from \citet{Bak&Svanholt1980}, \citet{Winnewisser1984} and \citet{Stolze1989} experimental works. 
We have identified a plethora of distinct $a$-type rotational transitions belonging to this species, which cover an energy level range between $J_\text{up} = 3$ up to $J_\text{up} = 11$ and correspond to different $K_\text{a} = 0, 1, 2$ ladders. This delineates nearly the entirety of the most prominent rotational spectroscopic features attributed to the $R$-branch $a$-type lines, rendering this detection exceptionally robust. Fig.~\ref{fig:H2CNCN_detected_lines} shows all of the transitions we have selected to perform the LTE fit for \ce{H2CNCN}, which are all those among targeted that are either unblended or present certain blending with other molecular species, but in which case the combined profiles closely match the observed spectrum. The rest of \ce{H2CNCN} lines covered by the observations are too faint to be detected or exhibit more significant blending with the emission from other molecular species (either unidentified or already modelled towards G+0.693), in all these cases being their predicted emission also consistent with the observed spectrum (see Fig.~\ref{fig:H2CNCN_blended_transitions} of Appendix~\ref{appendix:H2CNCN_blended_transitions}). The spectroscopic information of the transitions displayed in Fig.~\ref{fig:H2CNCN_detected_lines} is given in Table~\ref{tab:H2CNCN_spectroscopy}. All of them have been identified within a signal-to-noise ratio (S/N) in integrated intensity larger than 6 (see Table~\ref{tab:H2CNCN_spectroscopy}).

To achieve the best LTE modelling for \ce{H2CNCN}, we followed a two-step methodology. First, we fitted only the most unblended transitions (depicted in panels b), d), e) and j) of Fig.~\ref{fig:H2CNCN_detected_lines}), which provided the most accurate estimate for the $\text{FWHM}$. To do so, we ran AUTOFIT leaving the four parameters free and obtained a $\text{FWHM}$ of 21.9$\, \pm \,$1.4\kms, which we subsequently kept fixed in the second step. Then, we fitted all of the transitions shown in Fig.~\ref{fig:H2CNCN_detected_lines} and ran AUTOFIT again but now leaving only the other three parameters free. We obtained $v_\text{LSR} = 68.1 \pm 0.5$\kms, $T_\text{ex} = 7.9 \pm 0.3$\K and $N = (2.9 \pm 0.1)\times10^{12}\, \text{cm}^{-2}$ (see Table~\ref{tab:fitting_parameters}). The derived \ce{H2CNCN} column density translates into a total molecular abundance with respect to \ce{H2} of (2.1$\, \pm \,$0.3)$\times 10^{-11}$, assuming $N_\text{\ce{H2}} = 1.35\times10^{23} \, \text{cm}^{-2}$ \citep{Martin2008} with an associated uncertainty of $15\%$ to its value.

\begin{figure}[t!]
    \centering
    \includegraphics[width=\columnwidth]{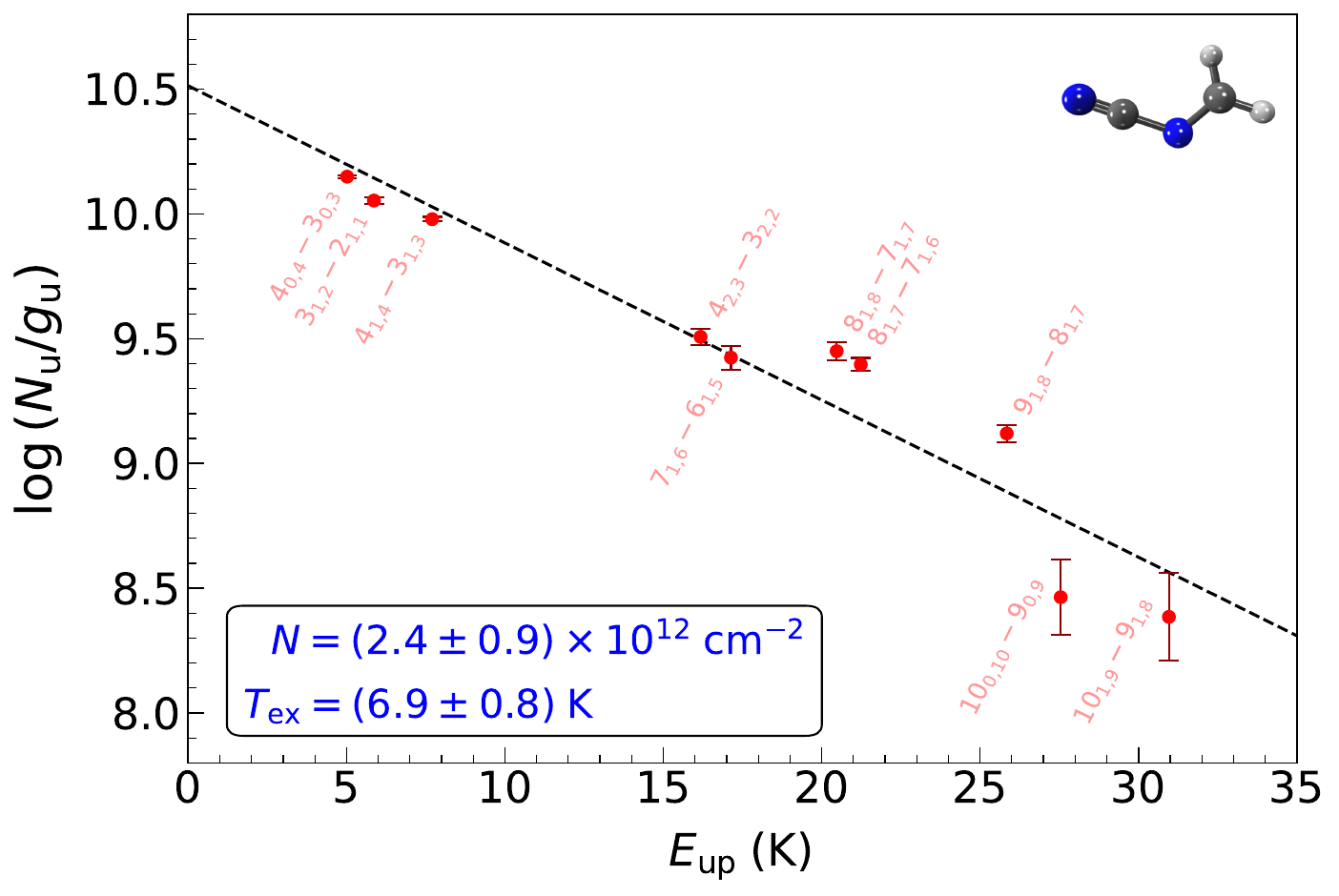}
    \caption{Rotational diagram of \ce{H2CNCN} using the unblended and partially blended transitions shown in Fig.~\ref{fig:H2CNCN_detected_lines} (red points and labels), but excluding those blended with species not yet identified (see Table~\ref{tab:H2CNCN_spectroscopy}). The dashed line depicts the best linear fit to the data, while the resulting values for the column density ($N$) and excitation temperature ($T_\text{ex}$) are displayed in blue within the black box. The molecular structure of \ce{H2CNCN} is shown in the upper right side.} 
    \label{fig:H2CNCN_rotational_diagram} 
\end{figure}

We have also performed a complementary rotational diagram analysis \citep{Goldsmith1999} for \ce{H2CNCN}, which is implemented in \textsc{MADCUBA} and incorporates two distinctive functionalities that provide a greater versatility in constrast with the method conventionally used. On one hand, \textsc{MADCUBA} rotational diagram accounts for opacity effects. Moreover, a very recent update allows to consider the predicted emission from all of the molecular species previously detected in the source. This is done by subtracting the predicted line profiles of the blending species from the observed data to generate an ``unblended'' spectral data set, from which the rotational diagram for the species of interest can be derived following the same procedure that would apply for unblended lines. 
Fig.~\ref{fig:H2CNCN_rotational_diagram} shows the rotational diagram we obtained for \ce{H2CNCN}. To compute it, we used all of the transitions shown in Fig.~\ref{fig:H2CNCN_detected_lines} but excluded those that exhibit blending with yet unidentified species (see Table~\ref{tab:H2CNCN_spectroscopy}), and employed their associated velocity integrated intensity over the line width as calculated by \textsc{MADCUBA}. This analysis yields $N = (2.4\pm0.9)\times10^{12}\, \text{cm}^{-2}$ and $T_\text{ex} = 6.9 \pm 0.8$\K, which are in perfect agreement with SLIM-AUTOFIT estimates.

\subsection{Analysis of $E$- and $Z$-\ce{HNCHCN}} \label{subsec:analysis_of_Z_E-HNCHCN}

\begin{figure*}[ht!]
    \centering
    \includegraphics[width=\textwidth]{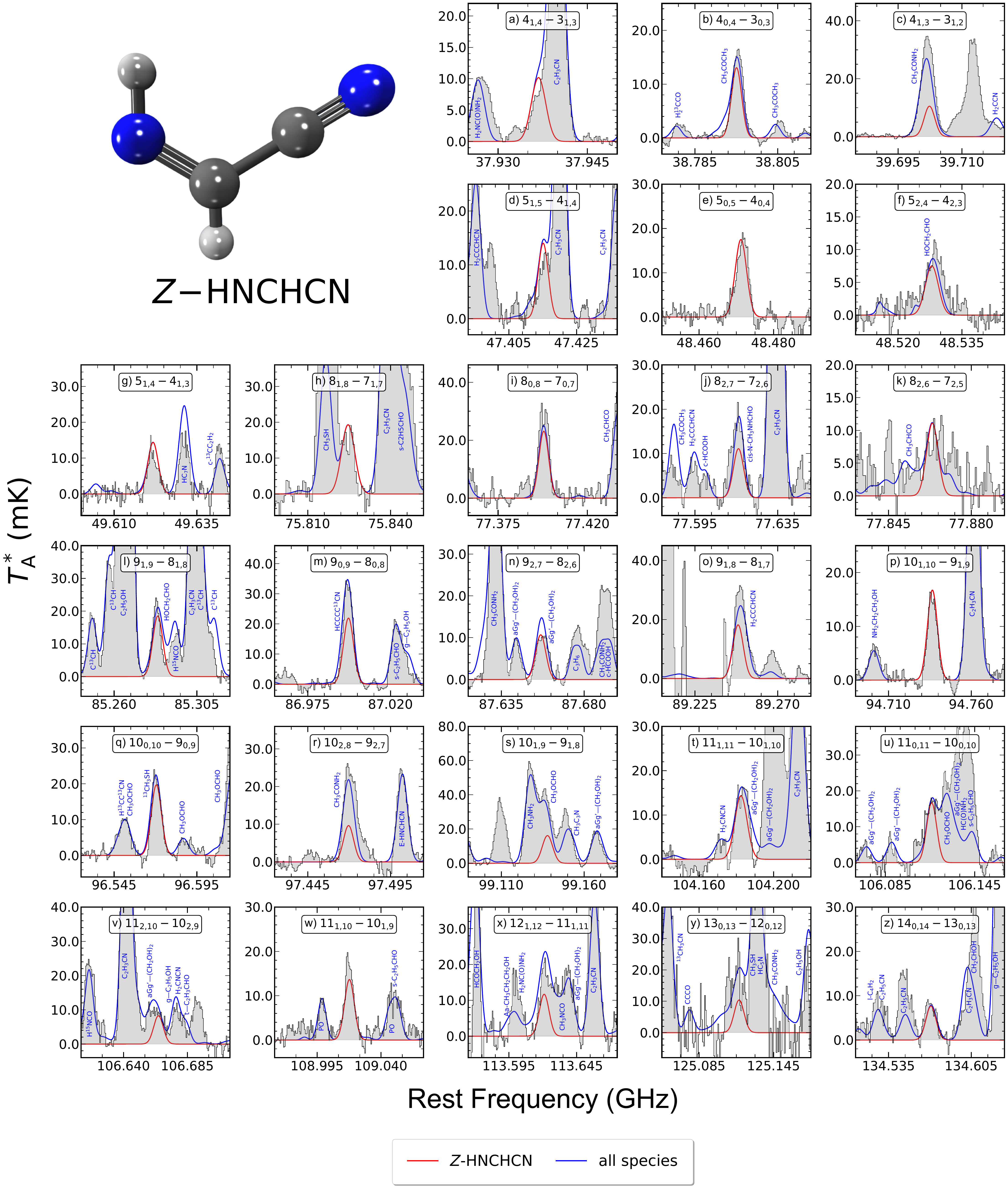}
    \caption{$Z$-\ce{HNCHCN} unblended or partially blended transitions detected towards G+0.693. The black histogram, grey-shaded areas, red and blue solid lines, blue labels and panel labels represent the same as indicated in Fig.~\ref{fig:H2CNCN_detected_lines}. Their spectroscopic information is given in Table~\ref{tab:Z,E-HNCHCN_spectroscopy} within Appendix~\ref{appendix:Z,E-HNCHCN_spectroscopy}. The molecular structure of $Z$-\ce{HNCHCN} is drawn in the upper left side, following the same color pattern as in Fig.~\ref{fig:H2CNCN_detected_lines}.} 
    \label{fig:Z-HNCHCN_detection} 
\end{figure*}

Both the $E$-,$Z$-isomers of the $C$-cyanomethanimine species (\ce{HNCHCN}) were previously detected towards G+0.693 by \citet{Rivilla2019}, albeit on the basis of a less sensitive IRAM 30 m spectral survey with a narrower spectral coverage (85$-$109\GHz). In this regard, the presence of new high-sensitivity Yebes 40 m observations at frequencies $\sim$30$-$50\GHz, along with the significant enhancement of the spectral sensitivity at also higher frequencies, has prompted us to revisit the detection of these isomers towards G+0.693. 

\begin{figure*}[ht!]
    \centering
    \includegraphics[width=\textwidth]{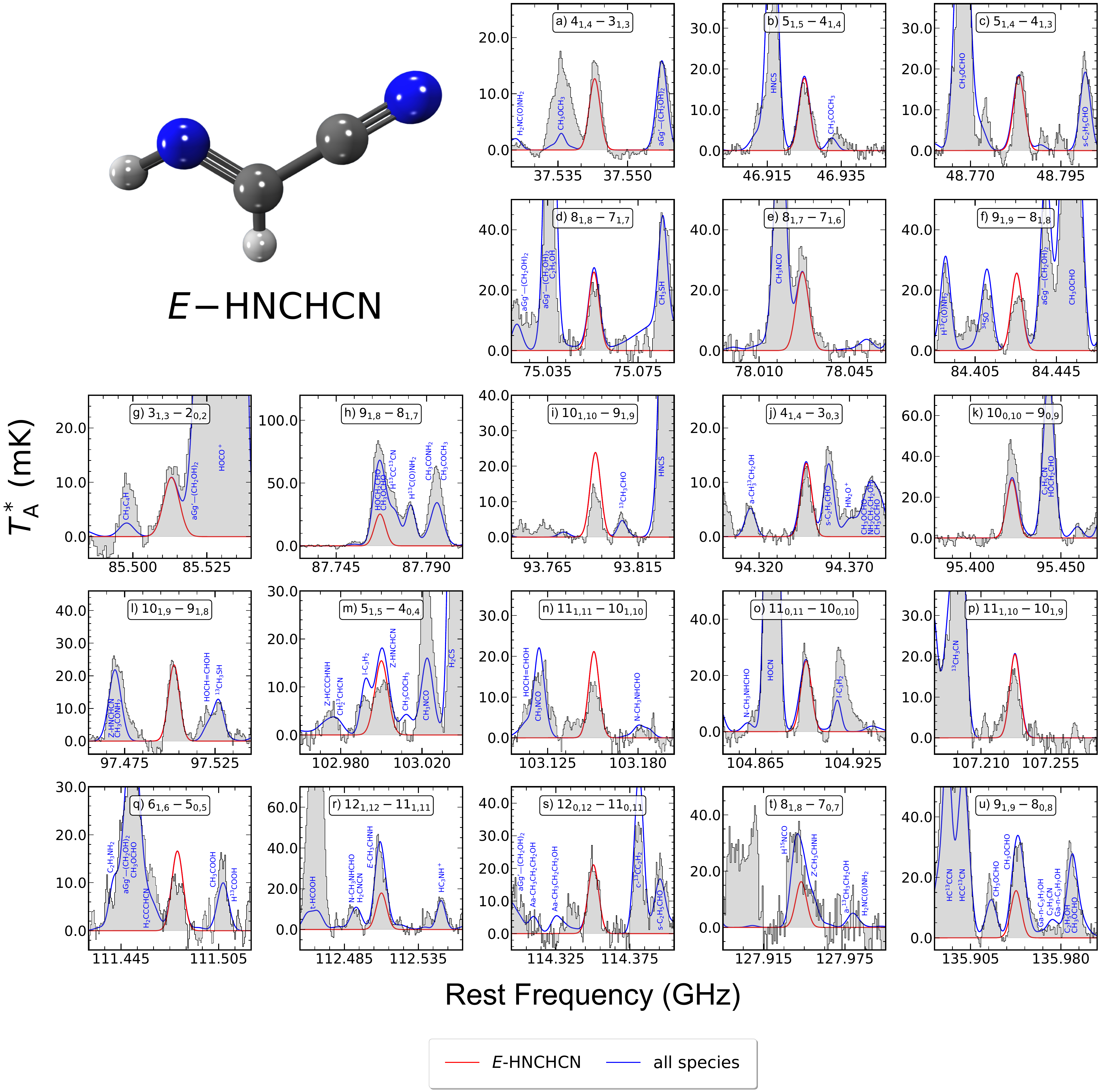}
    \caption{Same as Fig.~\ref{fig:Z-HNCHCN_detection} but for $E$-\ce{HNCHCN}. The spectroscopic information for all of these transitions is given in Table~\ref{tab:Z,E-HNCHCN_spectroscopy} within Appendix~\ref{appendix:Z,E-HNCHCN_spectroscopy}. $E$-\ce{HNCHCN} molecular structure is depicted in the upper left corner, colored according to the same palette as in Figs.~\ref{fig:H2CNCN_detected_lines} and \ref{fig:Z-HNCHCN_detection}.} 
    \label{fig:E-HNCHCN_detection} 
\end{figure*}

To perform the new analysis of these two species, we employed the same spectroscopic entries from the CDMS catalog that \citet{Rivilla2019} also used: 054512 for $Z$-\ce{HNCHCN} and 054513 for $E$-\ce{HNCHCN} (both dated November 2018), which contain the rotational transitions measured by \citet{Takano1990}, \citet{Zaleski2013} and \citet{Melosso2018}. We have almost tripled the number $Z$- and $E$-\ce{HNCHCN} unblended or partially blended lines detected towards G+0.693 with a S/N ratio in integrated intensity above 6, which are shown in Figs.~\ref{fig:Z-HNCHCN_detection} and \ref{fig:E-HNCHCN_detection}, respectively. For both isomers, these lines correspond to different $K_\text{a} = 0, 1, 2$ ladders of $a$-type rotational transitions sweeping a wider range of energy levels ($\sim$5$-$50\K) than those previously covered by \citet{Rivilla2019} ($\sim$21$-$34\K), and starting from $J_\text{up} = 4$ up to $J_\text{up} = 14$. Moreover, we have also targeted for the first time some $b$-type transitions belonging to $E$-\ce{HNCHCN} [see panels g), j), m), q), t) and u) within Fig.~\ref{fig:E-HNCHCN_detection}], facilitated by the relatively high $\mu_b$ dipole moment component, albeit smaller compared to its $\mu_a$ counterpart (2.51 D versus 3.25 D, respectively; \citealt{Takano1990}). The spectroscopic information of all these transitions is gathered in Table~\ref{tab:Z,E-HNCHCN_spectroscopy} within Appendix~\ref{appendix:Z,E-HNCHCN_spectroscopy}, which highlights how some of the lines that \citet{Rivilla2019} targeted as ``unblended'' actually exhibit slight blending from other species. Furthermore, we have now detected all of them with a nearly three times greater S/N ratio in integrated intensity.

We have carried out the LTE line fitting for the $Z$- and $E$-\ce{HNCHCN} isomers by using all of the transitions depicted in Figs.~\ref{fig:Z-HNCHCN_detection} and \ref{fig:E-HNCHCN_detection}, respectively. As already explained for \ce{H2CNCN} (see Sect.~\ref{subsec:detection_of_H2CNCN}), we followed a similar two-step approach in both cases, whose results are summarized in Table~\ref{tab:fitting_parameters}.
In the case of $Z$-\ce{HNCHCN}, we first constrained its $\text{FWHM}$ by fitting the unblended lines shown in panels e), g), p) and z) of Fig.~\ref{fig:Z-HNCHCN_detection} while leaving the four parameters unfixed, obtaining a $\text{FWHM}$ of 22.3$\, \pm \,$0.8\kms. As for $E$-\ce{HNCHCN}, we derived a similar $\text{FWHM}$ of 22.1$\, \pm \,$0.6\kms by fitting the seven unblended transitions displayed in panels b-d), j), l), o) and s) within Fig.~\ref{fig:E-HNCHCN_detection}. In the second step, we repeated the fit but including the rest of transitions shown in Fig.~\ref{fig:Z-HNCHCN_detection} for $Z$-\ce{HNCHCN} and Fig.~\ref{fig:E-HNCHCN_detection} for the $E$-isomer, while keeping the $\text{FWHM}$ fixed to the former values, respectively. We obtained $N = (7.5 \pm 0.3) \times 10^{13}\, \text{cm}^{-2}$, $T_\text{ex} = 14.1 \pm 0.8$\K, $v_\text{LSR} = 67.0 \pm 0.5$\kms for $Z$-\ce{HNCHCN}, whereas $N = (1.68 \pm 0.03) \times 10^{13}\, \text{cm}^{-2}$, $T_\text{ex} = 15.5 \pm 0.5$\K and $v_\text{LSR} = 67.8 \pm 0.2$\kms for $E$-\ce{HNCHCN}. The column density derived for each isomer results into a molecular abundance with respect to \ce{H2} of (5.5$\, \pm \,$0.9)$\times 10^{-10}$ for $Z$-\ce{HNCHCN} and of (1.2$\, \pm \,$0.2)$\times 10^{-10}$ for $E$-\ce{HNCHCN}, which were calculated following the same approach as explained in Sect.~\ref{subsec:detection_of_H2CNCN}.

As for $N$-cyanomethanimine, we have also performed a rotational diagram analysis for both the $Z$- and $E$-isomers of $C$-cyanomethanimine, which we show in Fig.~\ref{fig:Z,E-HNCHCN_rotational_diagram}. In line with the methodology we followed for \ce{H2CNCN} (see Sect.~\ref{subsec:detection_of_H2CNCN}), we constructed this diagram by using all of the unblended and partially blended rotational transitions which are shown in Figs.~\ref{fig:Z-HNCHCN_detection} and \ref{fig:E-HNCHCN_detection} for $Z$- and $E$-\ce{HNCHCN} respectively, while incorporating their associated velocity integrated intensity. As done for \ce{H2CNCN}, we did not include those transitions that are blended with species that have not yet been identified (see Table~\ref{tab:Z,E-HNCHCN_spectroscopy} within Appendix~\ref{appendix:Z,E-HNCHCN_spectroscopy}). The results derived from this complementary analysis are $N = (5.9\pm1.3)\times10^{13}\, \text{cm}^{-2}$ and $T_\text{ex} = 14.4 \pm 1.3$\K for $Z$-\ce{HNCHCN}, and $N = (1.3\pm0.3)\times10^{13}\, \text{cm}^{-2}$ and $T_\text{ex} = 16.5 \pm 1.9$\K for $E$-\ce{HNCHCN}, being in both cases consistent with SLIM-AUTOFIT outcomes.

The substantial increase in the number of detected lines, along with their improved signal-to-noise ratio and broader energy distribution, has resulted into a more accurate characterization of the physical parameters describing the emission from these two isomers. Our enhanced fit now sets a higher but consistently similar $T_\text{ex}$ for both isomers of almost twice the value derived by \citet{Rivilla2019} for $Z$-\ce{HNCHCN} (8$\, \pm \,$2\K), which was the only one they could explicitly obtain from the lower sensitivity data. The notable difference in the $T_\text{ex}$, linked to the more precise scrutiny of the molecular line blending from other species, leaves its imprint in the column density estimates as well, which are now set to be $\sim$2$-$3 times lower in comparison with \citet{Rivilla2019} findings. Similarly, we have also computed a rather lower $Z/E$-\ce{HNCHCN} abundance ratio towards G+0.693 of 4.5$\, \pm \,$0.2, which remains consistent within uncertainty with the former estimate (6.1$\, \pm \,$2.4). Nonetheless, we emphasise the comparably refined accuracy concerning this new result, which highlights the direct impact that a full spectroscopic coverage exerts on determining the physical parameters.

\subsection{The $C/N$-cyanomethanimine abundance ratio towards G+0.693}
\label{sec:the_C/N_abundance_ratio}

Based on the new column density estimates derived for both the $Z$- and $E$-\ce{HNCHCN} isomers (see Table~\ref{tab:fitting_parameters}), we have computed a total column density for $C$-cyanomethanimine towards G+0.693 of (9.2$\, \pm \,$0.3)$\times 10^{13} \, \text{cm}^{-2}$, which results in a total molecular abundance with respect to \ce{H2} of (6.8$\, \pm \,$1.0)$\times 10^{-10}$. This value, although $\sim$3 times lower than that previously derived by \citet{Rivilla2019}, is of the same order of the abundances of other complex nitriles detected towards this cloud, such as \ce{CH3CN}, \ce{C2H3CN} or \ce{C2H5CN} \citep{Zeng2018}. This indicates that $C$-cyanomethanimine might indeed be a relatively abundant species in the ISM, which could favor its detection towards other astronomical sources. 

Nevertheless, this scenario significantly shifts for the $N$-cyanomethanimine species (\ce{H2CNCN}), which exhibits a remarkably lower molecular abundance with respect to \ce{H2} of (2.1$\, \pm \,$0.3)$\times 10^{-11}$. In fact, \ce{H2CNCN} emerges as one of the least abundant species detected thus far towards G+0.693, being $\sim$2$-$3 times lower in abundance in comparison to other even more complex nitrogen-bearing species already identified in this cloud \citep{Rivilla2022c}. This disparity becomes even sharper with respect to both $C$-cyanomethanimine isomers, for which we have computed $Z,E$-\ce{HNCHCN}/\ce{H2CNCN} isomeric ratios of 25.9$\, \pm \,$1.6 and 5.8$\, \pm \,$0.3, respectively. Consequently, we have derived a $C/N$-cyanomethanimine abundance ratio of 31.8$\, \pm \,$1.8 towards G+0.693. 

At this point, it is worth mentioning that while the $Z$- and $E$-\ce{HNCHCN} isomers exhibit quite similar $T_\text{ex}$ of $\sim$14\K, we have derived a much lower value of $\sim$8\K for \ce{H2CNCN} (see Table~\ref{tab:fitting_parameters}). Although we have already demonstrated the reliability of our results regarding the LTE fitting of these species, we have also evaluated the possibility of a $T_\text{ex}$ of $\sim$14\K for \ce{H2CNCN}. In this scenario, we have obtained a poorer fit and a reduction of $\sim$25\% in its estimated column density, which results into an even lower molecular abundance and hence a greater $C/N$-cyanomethanimine ratio. In any case, our observational results set the $N$-isomer being over one order of magnitude less abundant compared to the more stable $C$-cyanomethanimine species, which suggests that its detection towards other sources could become a much more challenging task.

\begin{figure*}[ht!]
\gridline{\fig{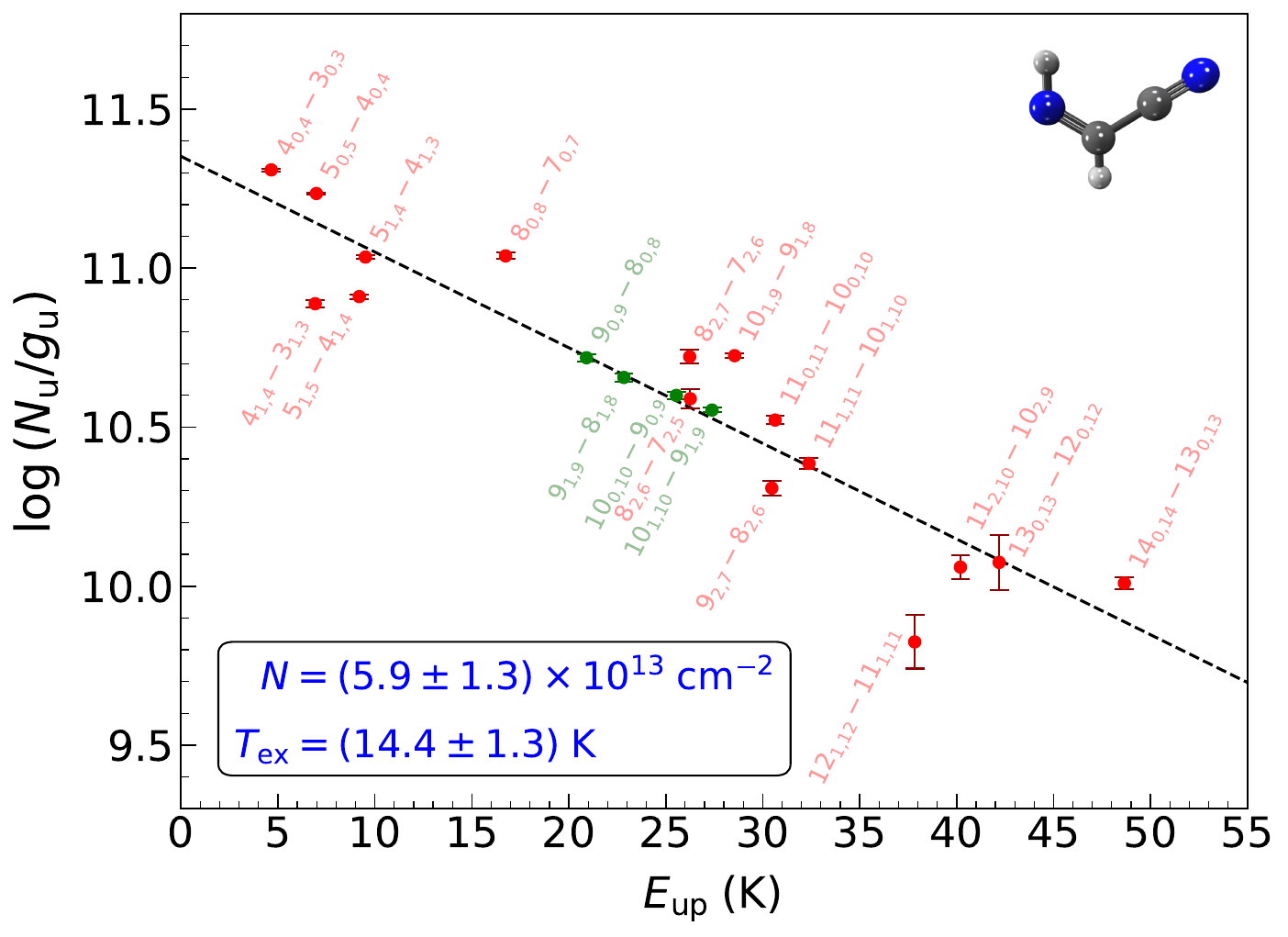}{0.5\textwidth}{\qquad\quad(a) $Z$-\ce{HNCHCN}}
          \fig{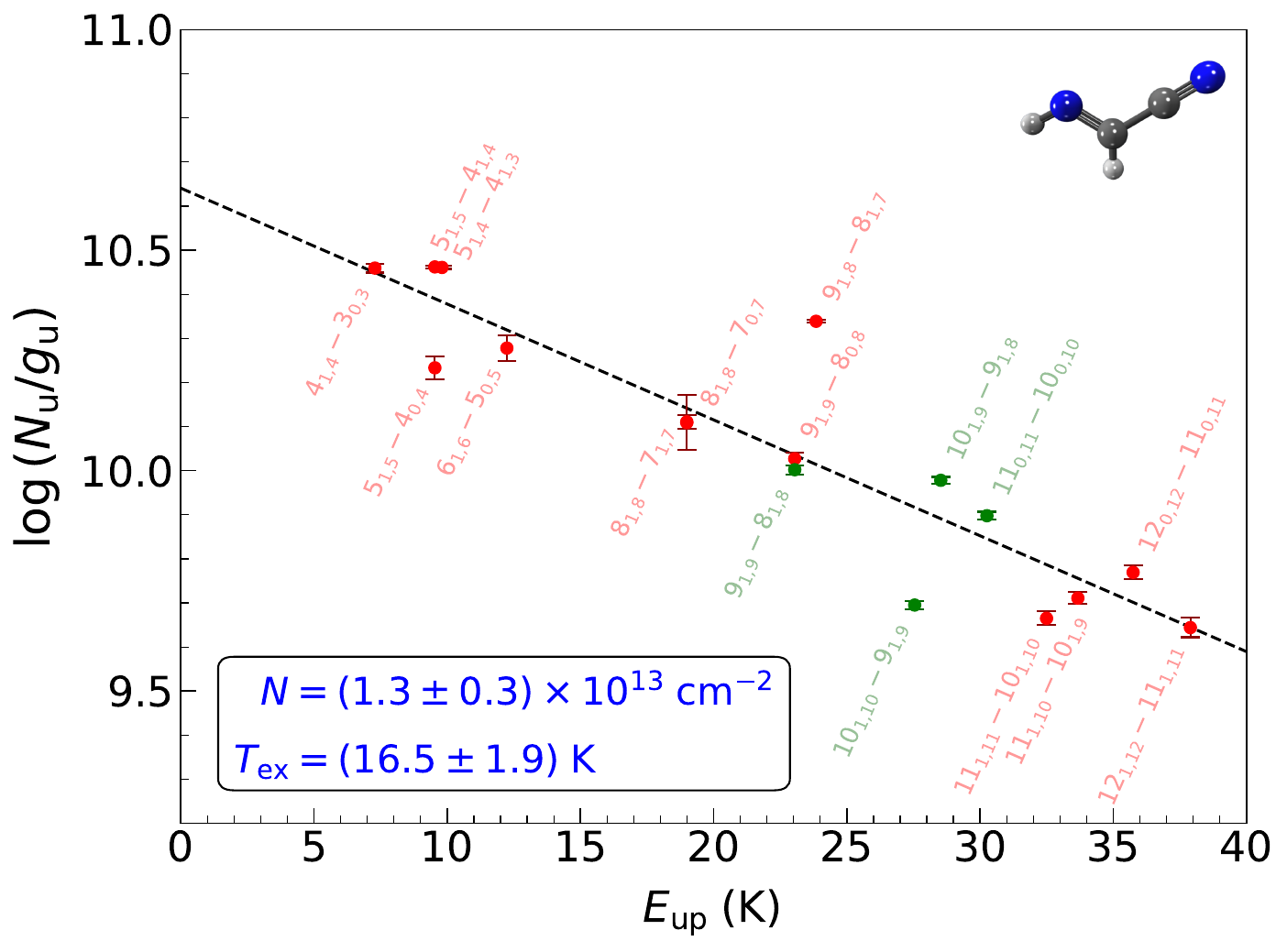}{0.5\textwidth}{\qquad\quad(b) $E$-\ce{HNCHCN}}
          }
\caption{Rotational diagrams of $Z$-\ce{HNCHCN} (left) and $E$-\ce{HNCHCN} (right) computed using these species unblended and partially blended transitions detected towards G+0.693 (colored points and labels, see Figs.~\ref{fig:Z-HNCHCN_detection} and \ref{fig:E-HNCHCN_detection} for each species, respectively), excluding those contaminated with yet unknown species (see Table~\ref{tab:Z,E-HNCHCN_spectroscopy} within Appendix~\ref{appendix:Z,E-HNCHCN_spectroscopy}). Transitions in green correspond to those previously targeted by \citet{Rivilla2019} that these authors used to perform the LTE fit of these species, whereas the rest of lines we have identified are displayed in red. The dashed line delineates the best linear fit to the data achieved. The resulting column density ($N$) and excitation temperature ($T_\text{ex}$) estimates are presented in blue framed within the black box. The molecular structure of each isomer is plotted in the upper right side.
\label{fig:Z,E-HNCHCN_rotational_diagram}}
\end{figure*}

%%%%%%%%%%%%%%%%%%%%%%%%%%%%%%%%%%%%%%%%%%%%%%%%%%%%%%%%%%
%%%%%%%%%%%%%%%%%%%%%%% DISCUSSION %%%%%%%%%%%%%%%%%%%%%%%
%%%%%%%%%%%%%%%%%%%%%%%%%%%%%%%%%%%%%%%%%%%%%%%%%%%%%%%%%%

\vspace{1cm}
\section{Discussion}
\label{sec:discussion}

\subsection{High-energy isomers in the ISM}\label{sec:discussion_high_energy_isomers_in_the_ISM}

In astrochemistry, there is still a prevalent belief that the search for high-energy isomers in the ISM is bound to failure, since these isomers are expected to exhibit molecular abundances which could be several orders of magnitude lower than their most stable analogues. This idea is rooted on the so-called minimum energy principle (MEP), which states a strong correlation between the observed abundances of the different members that compose a given isomeric family and their relative energies, with the most stable member expected to exhibit the highest abundance \citep{Lattelais2009}. In this respect, the cyanomethanimines family analysed in this work would seem to adhere to this principle. However, the detection of $N$-cyanomethanimine towards G+0.693 cast doubts on the reliability of the MEP as a predictive tool for assessing high-energy isomers traceability in the ISM, as discussed below.

According to the original formulation of the MEP by \citet{Lattelais2009}, the abundance ratios between structural isomers (denoting those molecular species sharing the same generic formula but whose constituent atoms are arranged differently) must be governed by thermodynamic equilibrium, scaling with $\exp{(-\Delta E/T_\text{kin})}$\footnote{Rigorously, the relative population under thermodynamic equilibrium should be described in terms of the Gibbs free energy associated with their isomerization reaction ($\Delta G$). However, the thermal correction of the free energy is not usually provided for many interstellar isomers, and in particular for $N$-cyanomethanimine. Therefore, to compare it with the isomeric ratios derived from the observations, we used the expression using their relative energies instead ($\Delta E$), which was demonstrated to be also reliable for the $C$-cyanomethanimine stereoisomers (\citealt{GarciadelaConcepcion2021}).}, where $\Delta E$ is their electronic energy difference and $T_\text{kin}$ indicates the kinetic temperature of the gas. This points to their relative abundances being determined by isomerization mechanisms, which would prevail over other processes. Nevertheless, this is far from being the case of $N$-cyanomethanimine. With an estimated $\Delta E$ of 3570$\, \pm \,$130\K \citep{Puzzarini2015} with respect to the most stable $Z$-\ce{HNCHCN} isomer, its abundance based on thermodynamical considerations is predicted to be more than 10 orders of magnitude lower in comparison to $C$-cyanomethanimine at the $T_\text{kin}$ of G+0.693 ($\sim$70$-$140; \citealt{Zeng2018}), nearly 9 orders of magnitude below to what has actually been observed. Therefore, it ought to be explained in terms of kinetics instead, being the most plausible explanation for this probably related to the very distinct molecular structures of the $N$- and $C$-cyanomethanimine species, which makes their unimolecular isomerization impossible under ISM conditions since it must involve a reaction intermediate far superior in energy to both species. 

On top of that, the $N$- and $C$-cyanomethanimine isomers do not emerge as the unique exception to the MEP, since recent detections in the ISM of many other isomeric families has also raised serious doubts about its general applicability regarding structural isomers. Some of these findings have demonstrated that high-energy structural isomers can share similar abundances to their more stable counterparts, as it is the case of the \ce{H2CN} and \ce{H2NC} isomers \citep{Cabezas2021, Agundez2023, SanAndres2023} and the two most stable isomers within the \ce{C3H4O} family ($trans$-propenal, \ce{CH2CHCHO}, and methyl ketene, \ce{CH3CHCO}; \citealt{Bermudez2018}; \citealt{Fuentetaja2023}), or even higher, as noted for the \ce{C3H2O}, \ce{C2H4O2} and \ce{C2H5N2O} isomeric families (see, e.g., \citealt{Loomis2015} and \citealt{Shingledecker2019}; \citealt{Mininni2020, Rivilla2023}, respectively). Therefore, it is clear that the MEP cannot be taken as a strict rule to argue about the actual detectability of high-energy structural isomers in the ISM.

\begin{figure}[t!]
    \centering
    \includegraphics[width=\columnwidth]{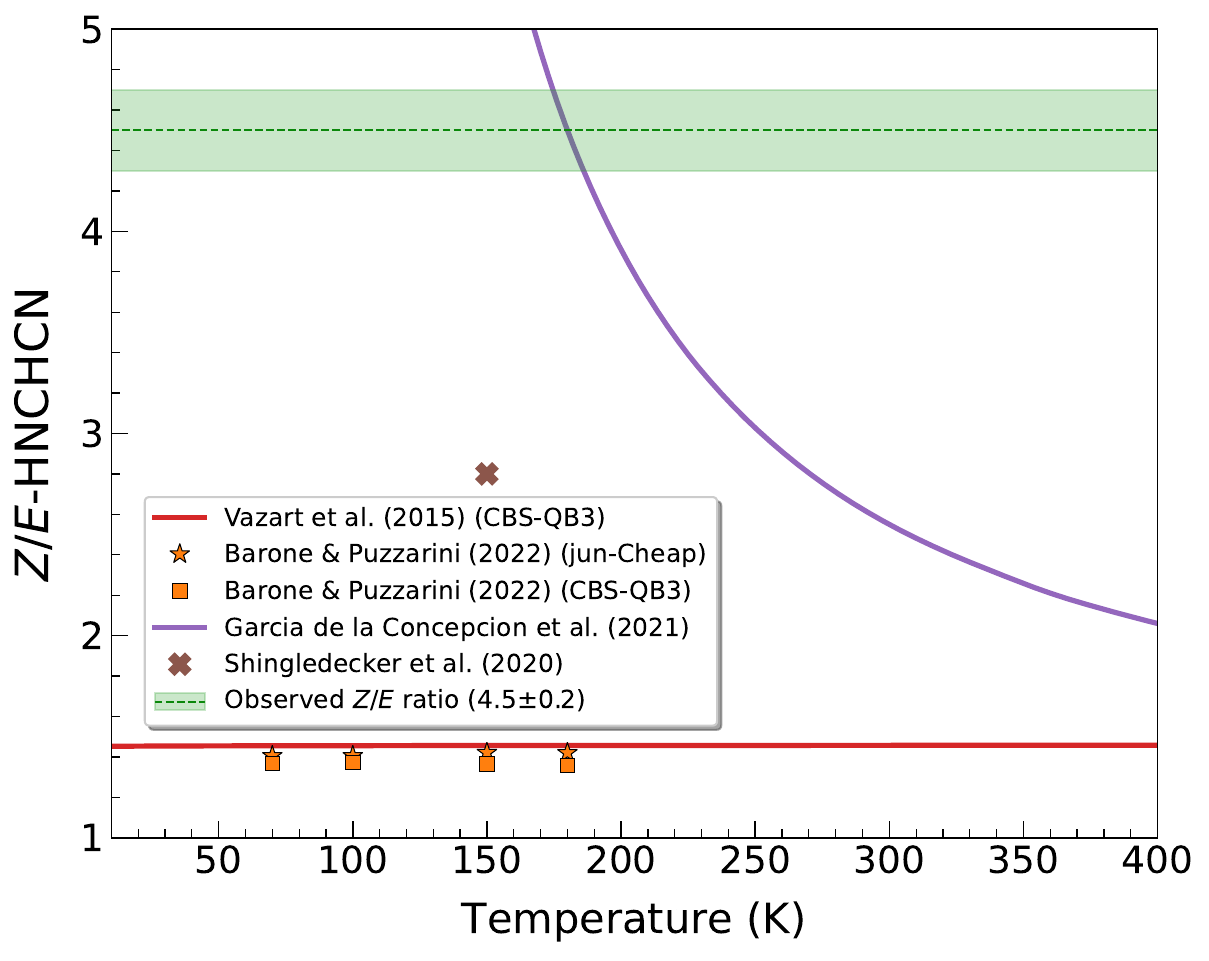} 
%    \caption{The $Z/E$-cyanomethanimine abundance ratio.} 
    \caption{The $Z/E$-\ce{HNCHCN} abundance ratio dependence on temperature as predicted by \citet{Vazart2015} and \citet{Barone&Puzzarini2022} chemical models for the \ce{CN + CH2NH} gas-phase reaction (red line and orange symbols, respectively), \citet{Shingledecker2020} astrochemical simulations (brown cross) and thermodynamics (purple line; \citealt{GarciadelaConcepcion2021}). The green dashed line and shaded area delineate the observed value encompassed by its 1$\sigma$ uncertainty (4.5$\, \pm \,$0.2). \citet{Barone&Puzzarini2022} estimates are retrieved for certain temperatures relevant to those derived for G+0.693 ($\sim$70$-$140\K; \citealt{Zeng2018}), as no analytical expression for the rate constant is provided.} 
    \label{fig:Z/E_abundance_ratio} 
\end{figure}

In this regard, the MEP only appears to primarily apply for some stereoisomers (also referred as spatial isomers, i.e., those species that possess identical constitutions but differ in the three dimensional orientation of their bonding atoms), where the $Z$- and $E$-\ce{HNCHCN} isomeric pair of $C$-cyanomethanimine emerge as the nearest example. In fact, the energy difference of these two isomers is 309$\, \pm \,$72\K \citep{Takano1990}, which based on the gas kinetic temperatures of the G+0.693 cloud of $\sim$70$-$140\K \citep{Zeng2018}, results in an abundance ratio fairly consistent with that observed by \citet{Rivilla2019} and reported in this work. Although the energy barrier associated to their isomerization is huge ($\sim$15.95\kK; \citealt{Takano1990}), \citet{GarciadelaConcepcion2021} demonstrated that this process can take place, even at the low temperatures of G+0.693, when quantum tunneling is considered using a small curvature approximation \citep{Skodje1981}, and predicted a $Z/E$-\ce{HNCHCN} abundance ratio at 150\K which mimicked \citet{Rivilla2019} observational value (6.1$\, \pm \,$2.4). Our revised estimate for this ratio of 4.5$\, \pm \,$0.2 would correspond to a $T_\text{kin}$ of $\sim$180\K (see Fig.~\ref{fig:Z/E_abundance_ratio}), slightly higher than those typically associated to this source, but still in good agreement with the thermodynamic equilibrium prediction. Furthermore, as Fig.~\ref{fig:Z/E_abundance_ratio} shows, this mechanism emerges as the only plausible explanation for this ratio, as none of the formerly suggested chemical pathways are capable of describing the observations \citep{Vazart2015, Shingledecker2020, Barone&Puzzarini2022}.

Besides both $C$-cyanomethanimine isomers, the list of stereoisomers that also support the MEP rule is rather scarce, where the $Ga$ and $Aa$ conformers of $n$-propanol ($n$-\ce{C3H7OH}; \citealt{Jimenez-Serra2020}), the $anti$ and $gauche$ conformers of ethyl formate (\ce{CH3CH2(O)CHO}; \citealt{Rivilla2017}), and the $cis$ and $trans$ conformers of thioformic acid (\ce{HC(O)SH}; \citealt{GarciadelaConcepcion2022}) are some of the clearest examples. However, there is also growing evidence that thermodynamics cannot account for the observed abundance ratios of many other stereoisomers, such as the $cis$-$cis$ and $cis$-$trans$ conformers of carbonic acid (\ce{HOCOOH}; \citealt{Sanz-Novo2023}), the $cis$ and $trans$ conformers of methyl formate (\ce{CH3OCHO}; \citealt{Neill2012}), and the $cis$ and $trans$ conformers of formic acid (\ce{HCOOH}; \citealt{GarciadelaConcepcion2022}). We note that all of the stereoisomers so far detected in the ISM whose abundance ratios match the thermodynamic prediction exhibit energy differences $\lesssim$500\K, which appears to be a doable energetic boundary for quantum tunneling to efficiently enable their isomerization reactions at the low temperatures of the ISM, as \citet{GarciadelaConcepcion2021} demonstrated for several imines. On the other hand, the stereoisomers mentioned above that do not follow the thermodynamic ratio exhibit much higher energy differences, which surely prevents their direct isomerization and hence makes their abundance ratios to be shaped by chemical kinetics instead.

All in all, observational evidence increasingly indicates that the MEP lacks predictive capability regarding which species within a specific isomeric family are more readily detectable and, consequently, cannot be generally used as a reliable indicator of isomeric abundances. In this regard, the detection of $N$-cyanomethanimine towards G+0.693 provides robust evidence that high-energy isomers can also be found in the ISM. Therefore, although it is possible that in some cases the detection of such species would deserve deeper integrations, we note that it is equally important and necessary obtaining their spectroscopy to enable their interstellar identification, significantly contributing to provide a full inventory of the molecular complexity of the ISM.

The aforementioned results have also continued to support the prominent role of chemical kinetics, rather than thermodynamics, in establishing the observed abundance ratios between structural isomers. For this reason, we discuss in the following section the possible chemical pathways leading to \ce{H2CNCN}.

\subsection{\ce{H2CNCN} interstellar chemistry}
\label{sec:chemistry}

Over the last decade, several authors have investigated the chemistry of cyanomethanimines under interstellar conditions \citep{Vazart2015, Shivani2017, Shingledecker2020, Zhang2020}. However, biased by the exclusive detection in the ISM of the $Z$- and $E$-\ce{HNCHCN} isomers \citep{Zaleski2013, Rivilla2019}, many of these works mainly focused on the chemical processes involving these two species. Consequently, the potential chemical pathways yielding to \ce{H2CNCN} have often been overlooked, resulting in a limited characterization of the chemistry linked to this isomer.

\begin{figure}[t!]
    \centering
    \includegraphics[width=\columnwidth]{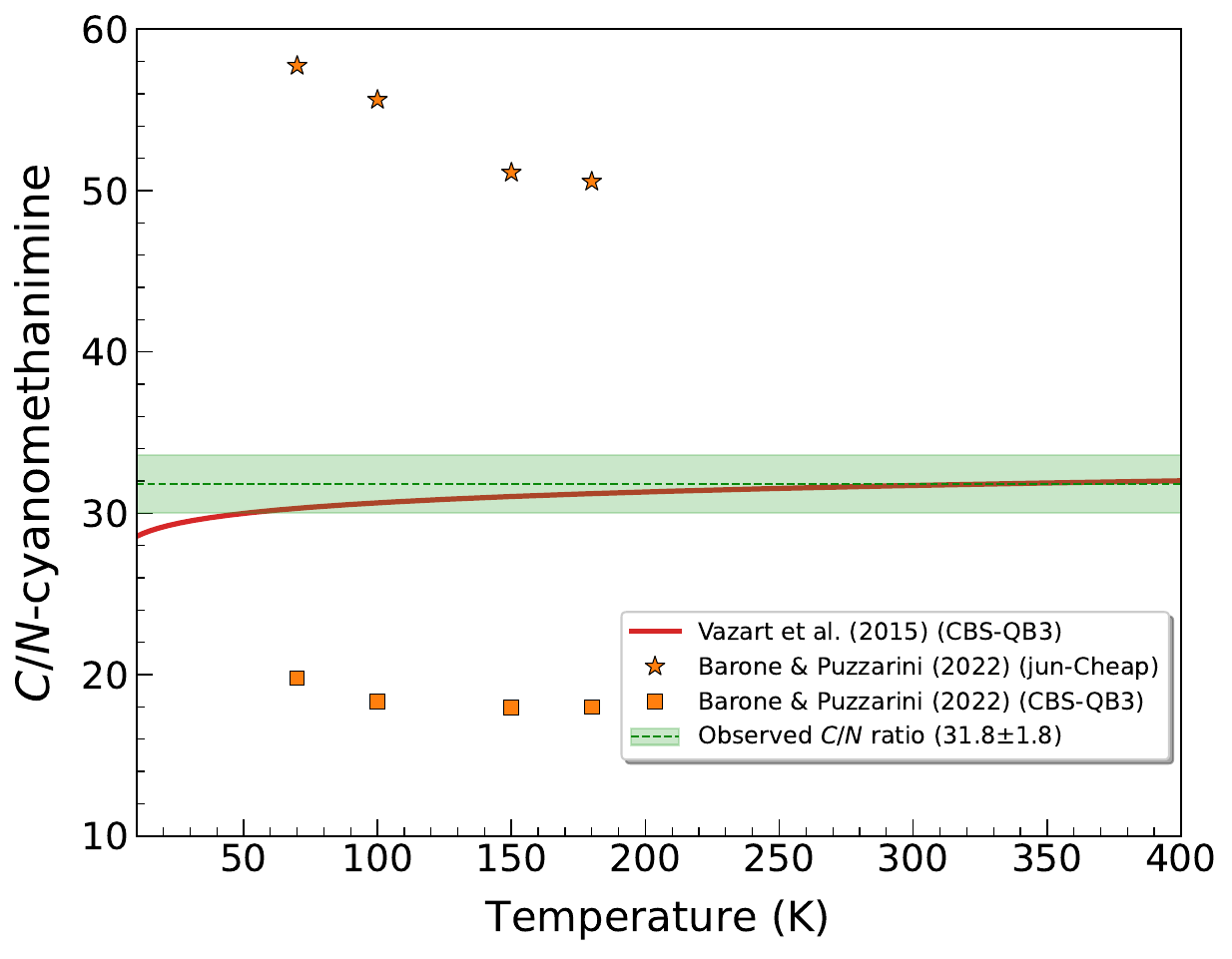}
    \caption{The $C/N$-cyanomethanimine abundance ratio dependence on temperature as predicted by \citet{Vazart2015} and \citet{Barone&Puzzarini2022} chemical models for the \ce{CN + CH2NH} reaction (red line and orange symbols, respectively). The green dashed line and shaded area delineate the observed value encompassed by its 1$\sigma$ uncertainty (31.8$\, \pm \,$1.8). \citet{Barone&Puzzarini2022} estimates are retrieved for certain temperatures relevant to those derived for G+0.693 ($\sim$70$-$140\K; \citealt{Zeng2018}), as no analytical expression for the rate constant is provided.} 
    \label{fig:C/N_abundance_ratio_reactions} 
\end{figure}

To date, there is just one formation mechanism leading to \ce{H2CNCN} that has been specifically studied. This route occurs through the gas-phase reaction between the cyanogen radical (\ce{CN}) and methanimine (\ce{CH2NH}), which \citet{Vazart2015} initially examined by performing quantum chemical calculations using a composite CBS-QB3 methodology. This reaction leads to the three cyanomethanimine isomers as the main products:

\begin{align}
    \ce{CN} + \ce{CH2NH} \; & \rightarrow \; Z,E\text{-}\ce{HNCHCN + H}, \tag{1.1} \label{reaction:CN+CH2CN->C-cyanomethanimine}\\
    {\color{white} \ce{CN} + \ce{CH2NH}} \; & \rightarrow \; \ce{H2CNCN + H}, \tag{1.2} \label{reaction:CN+CH2CN->N-cyanomethanimine}
\end{align}

\noindent
being the $C$-cyanomethanimine isomers formed when the \ce{CN} radical attacks the carbon atom of methanimine (route \ref{reaction:CN+CH2CN->C-cyanomethanimine}), while $N$-cyanomethanimine is produced from the attack on its N-side (route \ref{reaction:CN+CH2CN->N-cyanomethanimine}). All these processes are exothermic and proceed without an entrance barrier, making this reaction a feasible mechanism forming cyanomethanimines under interstellar conditions. Among the two possible pathways, the one leading to the $Z,E$-\ce{HNCHCN} isomers is much more thermodynamically favoured, hence arising as the predominant route. Although the kinetic calculations performed by these authors establish a rather low $Z/E$-\ce{HNCHCN} abundance ratio of $\sim$1.5 from this reaction (which is far below the 4.5$\, \pm \,$0.2 ratio we observed, see Fig.~\ref{fig:Z/E_abundance_ratio}), the prediction for the $C/N$-cyanomethanimine ratio fits extremely well to the observational value (see Fig.~\ref{fig:C/N_abundance_ratio_reactions}). Given that this ratio ought to be described in terms of chemical kinetics (see Sect.~\ref{sec:discussion_high_energy_isomers_in_the_ISM}), this result would point to the \ce{CN + CH2NH} reaction being mainly responsible for $C$- and $N$-cyanomethanimine chemistry in G+0.693. Moreover, its viability is also supported by the high abundance of the parent species (\ce{CH2NH} and \ce{CN}) found in this region (see \citealt{Zeng2018} and \citealt{Rivilla2019}, respectively). On top of that, this reaction can also proceed at a very fast pace. With associated rate constants at 150\K of $\sim$10$^{-11}-$10$^{-10} \; \text{cm}^{-3}\,\text{s}^{-1}$ as derived by \citet{Vazart2015}, the observed abundances of the cyanomethanimines would be reproduced through this reaction at timescales smaller than those related to depletion on dust grains due to the cooling of the gas ($\gtrsim$10$^5 \, \text{yr}$; \citealt{Requena-Torres2006}), even in spite of the low \ce{H2} densities of this cloud ($\sim$10$^4$$-$10$^5 \, \text{cm}^{-3}$; \citealt{Zeng2020}). 

Nonetheless, a new theoretical study recently carried out by \citet{Puzzarini&Barone2020} has revealed that the energy of the transition states along the reaction profile calculated by \citet{Vazart2015} could be slightly underestimated, causing their computed branching ratios not to be entirely accurate. In a subsequent study, \citet{Barone&Puzzarini2022} performed improved kinetic calculations using both the CBS-QB3 model as well as the more optimized and recently developed jun-Cheap scheme, outlining a slightly divergent $C/N$-cyanomethanimine abundance ratio as a function of the temperature with respect to \citet{Vazart2015} results, as shown in Fig.~\ref{fig:C/N_abundance_ratio_reactions}. The prediction for the $Z/E$-\ce{HNCHCN} ratio remains unchanged (see Fig.~\ref{fig:Z/E_abundance_ratio}). Nevertheless, the differences encountered in the predicted $C/N$-cyanomethanimine ratio are merely within a factor of $\sim$2 with respect to our observations, so that the gas-phase \ce{CN + CH2NH} reaction is still considered as the main mechanism likely producing cyanomethanimines in G+0.693. Moreover, its rate constant peaks at $\sim$150\K as derived by \citet{Barone&Puzzarini2022}, in line with the kinetic temperature of this cloud ($\sim$70$-$140\K; \citealt{Zeng2018}). Nonetheless, a more refined study of this process is needed to clarify the small discrepancies detected between the quantum chemical calculations and the astronomical observations.

Besides gas-phase processes, chemical reactions occurring on the icy mantles of interstellar dust grains could also be an important source producing \ce{H2CNCN} in G+0.693, where grain surface chemistry is believed to play a key role \citep{Rivilla2020a, Rivilla2021a, Rivilla2022b, Rivilla2023, Molpeceres2021, SanAndres2023}. However, little research has been done on these process for cyanomethanimine isomers, and especially for \ce{H2CNCN}. According to recent astrochemical models run by \citet{Zhang2020}, the surface analog of the already characterized gas-phase \ce{CN + CH2NH} reaction could be a second relevant route forming the three cyanomethanimine isomers. However, while it is well established that methanimine (\ce{CH2NH}) is mainly produced on grains making this reaction feasible \citep{Theule2011, Suzuki2016}, the hinted extraordinary reactivity that the \ce{CN} radical exhibits upon contact with \ce{H2O} molecules \citep{Rimola2018}, one of the major constituents of dust grain ices, would pose a significant barrier to its occurrence. Indeed, as it is also shown in the models of \citet{Zhang2020}, the contribution from this reaction to the total abundances of cyanomethanimines seems not to be as significant as that of its gas-phase counterpart, which can easily proceed once \ce{CH2NH} is desorbed into the gas. Furthermore, the observed abundances for the three cyanomethanimine isomers are reproduced within their models at timescales $<$10$^5 \, \text{yr}$ attending to the \ce{CN + CH2NH} gas-phase reaction, which strongly supports it as the primary and most efficient formation mechanism. However, the $C/N$-cyanomethanimine abundance ratio is not consistently explained according to their chemical network, indicating that other chemical processes might also need to be invoked. Moreover, the absence of comprehensive theoretical or experimental investigations examining this particular mechanism on grains forces us to assume the same branching ratio towards the formation of the three isomers in chemical models. Nonetheless, in spite of these shortcomings, it appears more likely that the role of the \ce{CN + CH2NH} reaction in producing cyanomethanimines is only relevant in the gas-phase.

Another possible surface pathway has been proposed by \citet{Vasconcelos2020}, whose experiments irradiating a \ce{N2}-\ce{CH4} icy mixture with cosmic rays produced the three isomers of cyanomethanimine. However, it remains unclear whether the presence of additional molecules in the ice, such as \ce{H2O}, \ce{CO} or \ce{CO2}, can alter the chemistry observed in these irradiation experiments. Intuitively, and because \ce{H2O}, \ce{CO} and \ce{CO2}  are the main constituents of interstellar ices (see, e.g., \citealt{Boogert2015, McClure2023}), the free path of the radicals derived from \ce{CH4} and \ce{N2} radiolysis is likely to encounter reaction partners of this ternary, reducing the amount of cyanomethanimines being produced and hence rendering this specific route probably not dominant.

\citet{Shivani2017} also proposed that both the $Z$- and $E$-\ce{HNCHCN} isomers could be formed in the grains through two consecutive hydrogenations of the cyanogen species (\ce{NCCN}), which is expected to be abundant in G+0.693 \citep{Rivilla2019}. However, the astrochemical model performed later by \citet{Shingledecker2020} showed that this route has a minor impact in the formation of $C$-cyanomethanimine in G+0.693. By analogy, it could be proposed that the hydrogenation of the high-energy metastable isomer of \ce{NCCN}, the isocyanogen species (\ce{CNCN}), could be another possible route remaining to be added in \ce{H2CNCN} solid-phase chemical network. Nonetheless, this alternative route is currently hampered by the poor knowledge on how this precursor could be synthesised through interstellar chemistry, for which the only pathway so far studied (\ce{CN + HNC} $\rightarrow$ \ce{CNCN + H}) has been shown to have high energy barriers when leading to \ce{CNCN} \citep{Petrie2004}. Furthermore, this species exhibits a low abundance in the ISM ($<$10$^{-10}$ with respect to \ce{H2}), with only one reported detection so far \citep{Agundez2018}. Indeed, using our spectral survey and assuming a $T_\text{ex}$ of 10\K and a $\text{FWHM}$ of 22.0\kms, we have just derived an upper limit for its molecular abundance towards G+0.693 of $\leq$2$\times$10$^{-11}$, which is lower than the abundance of \ce{H2CNCN} itself. Consequently, all evidence points to this route not being a promising alternative for boosting \ce{H2CNCN} abundance, as happened for the $Z$- and $E$-\ce{HNCHCN} isomers through the analogue reaction triggered by \ce{NCCN}. 

In summary, none of the chemical pathways proposed to date for grain surface chemistry seem capable of substantially increasing the abundance of the three cyanomethanimine isomers to adequately match the observational values. In this context, the gas-phase \ce{CN + CH2NH} reaction stands out as the primary formation route for these species, while the role of surface chemistry would primarily point to enhance the abundance of \ce{CH2NH} in the gas after ejection, thereby enabling the occurrence of this reaction.

\section{Summary and conclusions}
\label{sec:summary_and_conclusions}

We have presented the first detection in the ISM of $N$-cyanomethanimine (\ce{H2CNCN}), a highly significant species in the prebiotic context as a potential precursor of adenine, one of the fundamental nucleobases constituting RNA and DNA. We have been able to univocally identify more than a dozen of different $a$-type rotational transitions belonging to this species towards the Galactic Center G+0.693-0.027 molecular cloud, through the recently improved ultra-high-sensitivity spectral survey of this source using the Yebes 40 m and IRAM 30 m radio telescopes.  We have performed a LTE fit to the observed data, deriving a total column density of (2.9$\, \pm \,$0.1)$\times 10^{12} \; \text{cm}^{-2}$ which translates into a molecular abundance with respect to \ce{H2} of (2.1$\, \pm \,$0.3)$\times 10^{-11}$. This makes it one of the least abundant complex organic molecules so far detected towards this cloud, which demonstrates how the growing efforts in achieving a greater sensitivity on the observational data are pushing the limits of molecular species detectability in space.

The identification of $N$-cyanomethanimine in G+0.693-0.027 adds to the previous detection of its two more stable isomers in this same region: the $Z$- and $E$- isomers of $C$-cyanomethanimine (\ce{HNCHCN}). The expanded spectral coverage provided by the new observations to frequencies below 50\GHz allowed us to reevaluate the identification of these two stereoisomers as well. In both cases, we have almost tripled the number of transitions detected, encompassing a significantly broader range of energy levels. This has led to a more accurate characterization of the physical parameters tracing both isomers emission, which has resulted in a three times lower total column density for $C$-cyanomethanimine of (6.8$\, \pm \,$1.0)$\times 10^{-10}$ in comparison to that previously inferred. We have found the $C/N$-cyanomethanimine abundance ratio to be 31.8$\, \pm \,$1.8, which points to $N$-cyanomethanimine being over one order of magnitude less abundant compared to the more stable $C$-cyanomethanimine species. We have computed a $Z/E$-\ce{HNCHCN} isomeric ratio of 4.5$\, \pm \,$0.2 (consistent within uncertainty with the previous estimate of 6.1$\, \pm \,$2.4), and derived $Z,E$-\ce{HNCHCN}/\ce{H2CNCN} ratios of 25.9$\, \pm \,$1.6 and 5.8$\, \pm \,$0.3, respectively. The relative abundance between the $Z$- and $E$-\ce{HNCHCN} isomers closely aligns with the prediction based on thermodynamic equilibrium at the kinetic temperature of G+0.693 ($\sim$70$-$140\K). However, it should also be noted that thermodynamic control between both $C$-cyanomethanimine stereoisomers would only be established if their inter-conversion timescales are shorter than other competitive mechanisms (such as destruction processes), a task pending to be carefully checked in the following mechanistic studies related to these species. On the other hand, the $C/N$-cyanomethanimine ratio should be described in terms of the chemical reactions involving these species, since the isomerization of $N$-cyanomethanimine from $C$-cyanomethanimine seems to be unattainable at low temperatures. The three cyanomethanimine isomers are mainly formed in the gas through the reaction between methanimine (\ce{CH2NH}) and the cyanogen radical (\ce{CN}), which seems to be responsible for the observed $C/N$-cyanomethanimine abundance ratio.

\acknowledgments

We are very grateful to the Yebes 40 m and IRAM 30 m telescope staff for their precious help during the different observing runs. The 40 m radio telescope at Yebes Observatory is operated by the Spanish Geographic Institute (IGN; Ministerio de Transportes, Movilidad y Agenda Urbana). IRAM is supported by INSU/CNRS (France), MPG (Germany) and IGN (Spain). D.S.A., V.M.R and A.L.-G. acknowledge the funds provided by the Consejo Superior de Investigaciones Cient{\'i}ficas (CSIC) and the Centro de Astrobiolog{\'i}a (CAB) through the project 20225AT015 (Proyectos intramurales especiales del CSIC). D.S.A. also extends his gratitude for the financial support provided by the Comunidad de Madrid through the Grant PIPF-2022/TEC-25475. V.M.R. has also received support from the project RYC2020-029387-I funded by the Spanish Ministry of Science, Innovation and Universities/State Agency of Research MICIU/AEI/10.13039/501100011033 and by ``ESF, Investing in your future''.
I.J.-S., J.M.-P., L.C., A.M. and A.M.-H. acknowledge financial support through the Spanish grant PID2019-105552RB-C41 funded by MICIU/AEI/10.13039/501100011033. I.J.-S., J.M.-P., L.C., V.M.R., A.M. and A.M.-H. acknowledge also financial support through the Spanish grant PID2022-136814NB-I00 funded by MICIU/AEI/10.13039/501100011033 and by ``ERDF A way of making Europe''. A.M. has received support from grant PRE2019-091471 under project MDM-2017-0737-19-2 funded by MICIU/AEI/10.13039/501100011033, and by ``ERDF A way of making Europe''. A.M.-H. acknowledges funds from Grant MDM-2017-0737 Unidad de Excelencia ``Mar{\'i}a de Maeztu'' Centro de Astrobiolog{\'i}a (CAB, INTA-CSIC). M.S.N. acknowledges a Juan de la Cierva Postdoctoral Fellow proyect JDC2022-048934-I, funded by MICIU/AEI/10.13039/501100011033 and by the European Union ``NextGenerationEU''/PRTR.  B.T. and P.dV. thank the support from the Spanish MICIU through the project PID2019-107115GB-C21. B.T. also thanks the Spanish MICIU for funding support from grant PID2022-137980NB-I00. G.M. thanks the Japan Society for the Promotion of Science (JSPS International fellow P22013 and KAKENHI grant number JP22F22013).

%% To help institutions obtain information on the effectiveness of their 
%% telescopes the AAS Journals has created a group of keywords for telescope 
%% facilities.
%
%% Following the acknowledgments section, use the following syntax and the
%% \facility{} or \facilities{} macros to list the keywords of facilities used 
%% in the research for the paper.  Each keyword is check against the master 
%% list during copy editing.  Individual instruments can be provided in 
%% parentheses, after the keyword, but they are not verified.

\vspace{5mm}
%\facilities{HST(STIS), Swift(XRT and UVOT), AAVSO, CTIO:1.3m,
%CTIO:1.5m,CXO}

%% Similar to \facility{}, there is the optional \software command to allow 
%% authors a place to specify which programs were used during the creation of 
%% the manuscript. Authors should list each code and include either a
%% citation or url to the code inside ()s when available.

\software{Madrid Data Cube Analysis (\textsc{MADCUBA}) on ImageJ is a software developed at the Centre of Astrobiology (CAB) in Madrid, version 10.1.10 (2023 November 15) at \url{https://cab.inta-csic.es/madcuba} \citep{Martin2019}.}

%% Appendix material should be preceded with a single \appendix command.
%% There should be a \section command for each appendix. Mark appendix
%% subsections with the same markup you use in the main body of the paper.

%% Each Appendix (indicated with \section) will be lettered A, B, C, etc.
%% The equation counter will reset when it encounters the \appendix
%% command and will number appendix equations (A1), (A2), etc. The
%% Figure and Table counter will not reset.

\clearpage
\bibliography{biblio}{}

\begin{thebibliography}{}
\expandafter\ifx\csname natexlab\endcsname\relax\def\natexlab#1{#1}\fi
\providecommand{\url}[1]{\href{#1}{#1}}
\providecommand{\dodoi}[1]{doi:~\href{http://doi.org/#1}{\nolinkurl{#1}}}
\providecommand{\doeprint}[1]{\href{http://ascl.net/#1}{\nolinkurl{http://ascl.net/#1}}}
\providecommand{\doarXiv}[1]{\href{https://arxiv.org/abs/#1}{\nolinkurl{https://arxiv.org/abs/#1}}}

\bibitem[{{Ag{\'u}ndez} {et~al.}(2018){Ag{\'u}ndez}, {Marcelino}, \& {Cernicharo}}]{Agundez2018}
{Ag{\'u}ndez}, M., {Marcelino}, N., \& {Cernicharo}, J. 2018, \apjl, 861, L22, \dodoi{10.3847/2041-8213/aad089}

\bibitem[{{Ag{\'u}ndez} {et~al.}(2023){Ag{\'u}ndez}, {Roncero}, {Marcelino}, {Cabezas}, {Tercero}, \& {Cernicharo}}]{Agundez2023}
{Ag{\'u}ndez}, M., {Roncero}, O., {Marcelino}, N., {et~al.} 2023, \aap, 673, A24, \dodoi{10.1051/0004-6361/202346279}

\bibitem[{{Alberton} {et~al.}(2023){Alberton}, {Bizzocchi}, {Jiang}, {Melosso}, {Rivilla}, {Charmet}, {Giuliano}, {Caselli}, {Puzzarini}, {Alessandrini}, {Dore}, {Jim{\'e}nez-Serra}, \& {Mart{\'\i}n-Pintado}}]{Alberton2023}
{Alberton}, D., {Bizzocchi}, L., {Jiang}, N., {et~al.} 2023, \aap, 669, A93, \dodoi{10.1051/0004-6361/202244618}

\bibitem[{{Bak} {et~al.}(1978){Bak}, {Nielsen}, \& {Svanholt}}]{Bak1978}
{Bak}, B., {Nielsen}, O.~J., \& {Svanholt}, H. 1978, Chemical Physics Letters, 59, 330, \dodoi{10.1016/0009-2614(78)89106-4}

\bibitem[{{Bak} \& {Svanholt}(1980)}]{Bak&Svanholt1980}
{Bak}, B., \& {Svanholt}, H. 1980, Chemical Physics Letters, 75, 528, \dodoi{10.1016/0009-2614(80)80570-7}

\bibitem[{Balucani(2009)}]{Balucani2009}
Balucani, N. 2009, International Journal of Molecular Sciences, 10, 2304, \dodoi{10.3390/ijms10052304}

\bibitem[{{Barone} \& {Puzzarini}(2022)}]{Barone&Puzzarini2022}
{Barone}, V., \& {Puzzarini}, C. 2022, Frontiers in Astronomy and Space Sciences, 8, 255, \dodoi{10.3389/fspas.2021.814384}

\bibitem[{{Becker} {et~al.}(2016){Becker}, {Thoma}, {Deutsch}, {Gehrke}, {Mayer}, {Zipse}, \& {Carell}}]{Becker2016}
{Becker}, S., {Thoma}, I., {Deutsch}, A., {et~al.} 2016, Science, 352, 833, \dodoi{10.1126/science.aad2808}

\bibitem[{{Becker} {et~al.}(2019){Becker}, {Feldmann}, {Wiedemann}, {Okamura}, {Schneider}, {Iwan}, {Crisp}, {Rossa}, {Amatov}, \& {Carell}}]{Becker2019}
{Becker}, S., {Feldmann}, J., {Wiedemann}, S., {et~al.} 2019, Science, 366, 76, \dodoi{10.1126/science.aax2747}

\bibitem[{{Berm{\'u}dez} {et~al.}(2018){Berm{\'u}dez}, {Tercero}, {Motiyenko}, {Margul{\`e}s}, {Cernicharo}, {Ellinger}, \& {Guillemin}}]{Bermudez2018}
{Berm{\'u}dez}, C., {Tercero}, B., {Motiyenko}, R.~A., {et~al.} 2018, \aap, 619, A92, \dodoi{10.1051/0004-6361/201833267}

\bibitem[{{Bizzocchi} {et~al.}(2020){Bizzocchi}, {Prudenzano}, {Rivilla}, {Pietropolli-Charmet}, {Giuliano}, {Caselli}, {Mart{\'\i}n-Pintado}, {Jim{\'e}nez-Serra}, {Mart{\'\i}n}, {Requena-Torres}, {Rico-Villas}, {Zeng}, \& {Guillemin}}]{Bizzocchi2020}
{Bizzocchi}, L., {Prudenzano}, D., {Rivilla}, V.~M., {et~al.} 2020, \aap, 640, A98, \dodoi{10.1051/0004-6361/202038083}

\bibitem[{{Boogert} {et~al.}(2015){Boogert}, {Gerakines}, \& {Whittet}}]{Boogert2015}
{Boogert}, A.~C.~A., {Gerakines}, P.~A., \& {Whittet}, D. C.~B. 2015, \araa, 53, 541, \dodoi{10.1146/annurev-astro-082214-122348}

\bibitem[{{Br{\"u}nken} {et~al.}(2010){Br{\"u}nken}, {Belloche}, {Mart{\'\i}n}, {Verheyen}, \& {Menten}}]{Brunken2010}
{Br{\"u}nken}, S., {Belloche}, A., {Mart{\'\i}n}, S., {Verheyen}, L., \& {Menten}, K.~M. 2010, \aap, 516, A109, \dodoi{10.1051/0004-6361/200912456}

\bibitem[{{Cabezas} {et~al.}(2021){Cabezas}, {Ag{\'u}ndez}, {Marcelino}, {Tercero}, {Cuadrado}, \& {Cernicharo}}]{Cabezas2021}
{Cabezas}, C., {Ag{\'u}ndez}, M., {Marcelino}, N., {et~al.} 2021, \aap, 654, A45, \dodoi{10.1051/0004-6361/202141491}

\bibitem[{{Chakrabarti} \& {Chakrabarti}(2000)}]{Chakrabarti&Chakrabarti2000}
{Chakrabarti}, S., \& {Chakrabarti}, S.~K. 2000, \aap, 354, L6, \dodoi{10.48550/arXiv.astro-ph/0001079}

\bibitem[{{Choe}(2018)}]{Choe2018}
{Choe}, J.~C. 2018, Chemical Physics Letters, 708, 71, \dodoi{10.1016/j.cplett.2018.08.004}

\bibitem[{{Colzi} {et~al.}(2022){Colzi}, {Mart{\'i}n-Pintado}, {Rivilla}, {Jim{\'e}nez-Serra}, {Zeng}, {Rodr{\'i}guez-Almeida}, {Rico-Villas}, {Mart{\'i}n}, \& {Requena-Torres}}]{Colzi2022}
{Colzi}, L., {Mart{\'i}n-Pintado}, J., {Rivilla}, V.~M., {et~al.} 2022, \apjl, 926, L22, \dodoi{10.3847/2041-8213/ac52ac}

\bibitem[{{Endres} {et~al.}(2016){Endres}, {Schlemmer}, {Schilke}, {Stutzki}, \& {M{\"u}ller}}]{Endres2016}
{Endres}, C.~P., {Schlemmer}, S., {Schilke}, P., {Stutzki}, J., \& {M{\"u}ller}, H. S.~P. 2016, Journal of Molecular Spectroscopy, 327, 95, \dodoi{10.1016/j.jms.2016.03.005}

\bibitem[{Evans {et~al.}(1991)Evans, Lorencak, Ha, \& Wentrup}]{Evans1991}
Evans, R.~A., Lorencak, P., Ha, T.~K., \& Wentrup, C. 1991, Journal of the American Chemical Society, 113, 7261, \dodoi{10.1021/ja00019a026}

\bibitem[{{Fatima} {et~al.}(2023){Fatima}, {M{\"u}ller}, {Zingsheim}, {Lewen}, {Rivilla}, {Jim{\'e}nez-Serra}, {Mart{\'\i}n-Pintado}, \& {Schlemmer}}]{Fatima2023}
{Fatima}, M., {M{\"u}ller}, H. S.~P., {Zingsheim}, O., {et~al.} 2023, \aap, 680, A25, \dodoi{10.1051/0004-6361/202347112}

\bibitem[{Ferris \& Hagan(1984)}]{Ferris&Hagan1984}
Ferris, J.~P., \& Hagan, W.~J. 1984, Tetrahedron, 40, 1093, \dodoi{https://doi.org/10.1016/S0040-4020(01)99315-9}

\bibitem[{{Fuentetaja} {et~al.}(2023){Fuentetaja}, {Berm{\'u}dez}, {Cabezas}, {Ag{\'u}ndez}, {Tercero}, {Marcelino}, {Pardo}, {Margul{\`e}s}, {Motiyenko}, {Guillemin}, {de Vicente}, \& {Cernicharo}}]{Fuentetaja2023}
{Fuentetaja}, R., {Berm{\'u}dez}, C., {Cabezas}, C., {et~al.} 2023, \aap, 671, L6, \dodoi{10.1051/0004-6361/202245732}

\bibitem[{{Garc{\'i}a de la Concepci{\'o}n} {et~al.}(2021){Garc{\'i}a de la Concepci{\'o}n}, {Jim{\'e}nez-Serra}, {Carlos Corchado}, {Rivilla}, \& {Mart{\'i}n-Pintado}}]{GarciadelaConcepcion2021}
{Garc{\'i}a de la Concepci{\'o}n}, J., {Jim{\'e}nez-Serra}, I., {Carlos Corchado}, J., {Rivilla}, V.~M., \& {Mart{\'i}n-Pintado}, J. 2021, \apjl, 912, L6, \dodoi{10.3847/2041-8213/abf650}

\bibitem[{{Garc{\'\i}a de la Concepci{\'o}n} {et~al.}(2022){Garc{\'\i}a de la Concepci{\'o}n}, {Colzi}, {Jim{\'e}nez-Serra}, {Molpeceres}, {Corchado}, {Rivilla}, {Mart{\'\i}n-Pintado}, {Beltr{\'a}n}, \& {Mininni}}]{GarciadelaConcepcion2022}
{Garc{\'\i}a de la Concepci{\'o}n}, J., {Colzi}, L., {Jim{\'e}nez-Serra}, I., {et~al.} 2022, \aap, 658, A150, \dodoi{10.1051/0004-6361/202142287}

\bibitem[{{Gilbert}(1986)}]{Gilbert1986}
{Gilbert}, W. 1986, \nat, 319, 618, \dodoi{10.1038/319618a0}

\bibitem[{{Goldsmith} \& {Langer}(1999)}]{Goldsmith1999}
{Goldsmith}, P.~F., \& {Langer}, W.~D. 1999, \apj, 517, 209, \dodoi{10.1086/307195}

\bibitem[{Herbst {et~al.}(2020)Herbst, Vidali, \& Ceccarelli}]{Herbst2020}
Herbst, E., Vidali, G., \& Ceccarelli, C. 2020, ACS Earth and Space Chemistry, 4, 488, \dodoi{10.1021/acsearthspacechem.0c00043}

\bibitem[{{Jim{\'e}nez-Serra} {et~al.}(2020){Jim{\'e}nez-Serra}, {Mart{\'i}n-Pintado}, {Rivilla}, {Rodr{\'i}guez-Almeida}, {Alonso Alonso}, {Zeng}, {Cocinero}, {Mart{\'i}n}, {Requena-Torres}, {Mart{\'i}n-Domenech}, \& {Testi}}]{Jimenez-Serra2020}
{Jim{\'e}nez-Serra}, I., {Mart{\'i}n-Pintado}, J., {Rivilla}, V.~M., {et~al.} 2020, Astrobiology, 20, 1048, \dodoi{10.1089/ast.2019.2125}

\bibitem[{{Jim{\'e}nez-Serra} {et~al.}(2022){Jim{\'e}nez-Serra}, {Rodr{\'i}guez-Almeida}, {Mart{\'i}n-Pintado}, {Rivilla}, {Melosso}, {Zeng}, {Colzi}, {Kawashima}, {Hirota}, {Puzzarini}, {Tercero}, {de Vicente}, {Rico-Villas}, {Requena-Torres}, \& {Mart{\'i}n}}]{Jimenez-Serra2022}
{Jim{\'e}nez-Serra}, I., {Rodr{\'i}guez-Almeida}, L.~F., {Mart{\'i}n-Pintado}, J., {et~al.} 2022, \aap, 663, A181, \dodoi{10.1051/0004-6361/202142699}

\bibitem[{{Jones} {et~al.}(2012){Jones}, {Burton}, {Cunningham}, {Requena-Torres}, {Menten}, {Schilke}, {Belloche}, {Leurini}, {Mart{\'\i}n-Pintado}, {Ott}, \& {Walsh}}]{Jones2012}
{Jones}, P.~A., {Burton}, M.~G., {Cunningham}, M.~R., {et~al.} 2012, \mnras, 419, 2961, \dodoi{10.1111/j.1365-2966.2011.19941.x}

\bibitem[{{Jung} \& {Choe}(2013)}]{Jung&Choe2013}
{Jung}, S.~H., \& {Choe}, J.~C. 2013, Astrobiology, 13, 465, \dodoi{10.1089/ast.2013.0973}

\bibitem[{{Lattelais} {et~al.}(2009){Lattelais}, {Pauzat}, {Ellinger}, \& {Ceccarelli}}]{Lattelais2009}
{Lattelais}, M., {Pauzat}, F., {Ellinger}, Y., \& {Ceccarelli}, C. 2009, \apjl, 696, L133, \dodoi{10.1088/0004-637X/696/2/L133}

\bibitem[{{Li} {et~al.}(2020){Li}, {Wang}, {Qiao}, {Quan}, {Fang}, {Du}, {Li}, {Shen}, {Li}, {Li}, {Shi}, {Zhang}, \& {Zhang}}]{Li2020}
{Li}, J., {Wang}, J., {Qiao}, H., {et~al.} 2020, \mnras, 492, 556, \dodoi{10.1093/mnras/stz3337}

\bibitem[{{Loomis} {et~al.}(2015){Loomis}, {McGuire}, {Shingledecker}, {Johnson}, {Blair}, {Robertson}, \& {Remijan}}]{Loomis2015}
{Loomis}, R.~A., {McGuire}, B.~A., {Shingledecker}, C., {et~al.} 2015, \apj, 799, 34, \dodoi{10.1088/0004-637X/799/1/34}

\bibitem[{{Mart{\'i}n} {et~al.}(2019){Mart{\'i}n}, {Mart{\'i}n-Pintado}, {Blanco-S{\'a}nchez}, {Rivilla}, {Rodr{\'i}guez-Franco}, \& {Rico-Villas}}]{Martin2019}
{Mart{\'i}n}, S., {Mart{\'i}n-Pintado}, J., {Blanco-S{\'a}nchez}, C., {et~al.} 2019, \aap, 631, A159, \dodoi{10.1051/0004-6361/201936144}

\bibitem[{{Mart{\'i}n} {et~al.}(2008){Mart{\'i}n}, {Requena-Torres}, {Mart{\'i}n-Pintado}, \& {Mauersberger}}]{Martin2008}
{Mart{\'i}n}, S., {Requena-Torres}, M.~A., {Mart{\'i}n-Pintado}, J., \& {Mauersberger}, R. 2008, \apj, 678, 245, \dodoi{10.1086/533409}

\bibitem[{{Massalkhi} {et~al.}(2023){Massalkhi}, {Jim{\'e}nez-Serra}, {Mart{\'\i}n-Pintado}, {Rivilla}, {Colzi}, {Zeng}, {Mart{\'\i}n}, {Tercero}, {de Vicente}, \& {Requena-Torres}}]{Massalkhi2023}
{Massalkhi}, S., {Jim{\'e}nez-Serra}, I., {Mart{\'\i}n-Pintado}, J., {et~al.} 2023, \aap, 678, A45, \dodoi{10.1051/0004-6361/202346822}

\bibitem[{{McClure} {et~al.}(2023){McClure}, {Rocha}, {Pontoppidan}, {Crouzet}, {Chu}, {Dartois}, {Lamberts}, {Noble}, {Pendleton}, {Perotti}, {Qasim}, {Rachid}, {Smith}, {Sun}, {Beck}, {Boogert}, {Brown}, {Caselli}, {Charnley}, {Cuppen}, {Dickinson}, {Drozdovskaya}, {Egami}, {Erkal}, {Fraser}, {Garrod}, {Harsono}, {Ioppolo}, {Jim{\'e}nez-Serra}, {Jin}, {J{\o}rgensen}, {Kristensen}, {Lis}, {McCoustra}, {McGuire}, {Melnick}, {{\~A}-berg}, {Palumbo}, {Shimonishi}, {Sturm}, {van Dishoeck}, \& {Linnartz}}]{McClure2023}
{McClure}, M.~K., {Rocha}, W.~R.~M., {Pontoppidan}, K.~M., {et~al.} 2023, Nature Astronomy, 7, 431, \dodoi{10.1038/s41550-022-01875-w}

\bibitem[{{McGuire} {et~al.}(2012){McGuire}, {Loomis}, {Charness}, {Corby}, {Blake}, {Hollis}, {Lovas}, {Jewell}, \& {Remijan}}]{McGuire2012}
{McGuire}, B.~A., {Loomis}, R.~A., {Charness}, C.~M., {et~al.} 2012, \apjl, 758, L33, \dodoi{10.1088/2041-8205/758/2/L33}

\bibitem[{{Melosso} {et~al.}(2018){Melosso}, {Melli}, {Puzzarini}, {Codella}, {Spada}, {Dore}, {Degli Esposti}, {Lefloch}, {Bachiller}, {Ceccarelli}, {Cernicharo}, \& {Barone}}]{Melosso2018}
{Melosso}, M., {Melli}, A., {Puzzarini}, C., {et~al.} 2018, \aap, 609, A121, \dodoi{10.1051/0004-6361/201731972}

\bibitem[{Menor~Salván {et~al.}(2020)Menor~Salván, Bouza, Fialho, Burcar, Fernández, \& Hud}]{Menor-Salvan2020}
Menor~Salván, C., Bouza, M., Fialho, D.~M., {et~al.} 2020, ChemBioChem, 21, 3504, \dodoi{https://doi.org/10.1002/cbic.202000510}

\bibitem[{{Mininni} {et~al.}(2020){Mininni}, {Beltr{\'a}n}, {Rivilla}, {S{\'a}nchez-Monge}, {Fontani}, {M{\"o}ller}, {Cesaroni}, {Schilke}, {Viti}, {Jim{\'e}nez-Serra}, {Colzi}, {Lorenzani}, \& {Testi}}]{Mininni2020}
{Mininni}, C., {Beltr{\'a}n}, M.~T., {Rivilla}, V.~M., {et~al.} 2020, \aap, 644, A84, \dodoi{10.1051/0004-6361/202038966}

\bibitem[{{Molpeceres} {et~al.}(2021){Molpeceres}, {Garc{\'\i}a de la Concepci{\'o}n}, \& {Jim{\'e}nez-Serra}}]{Molpeceres2021}
{Molpeceres}, G., {Garc{\'\i}a de la Concepci{\'o}n}, J., \& {Jim{\'e}nez-Serra}, I. 2021, \apj, 923, 159, \dodoi{10.3847/1538-4357/ac2ebc}

\bibitem[{{Neill} {et~al.}(2012){Neill}, {Muckle}, {Zaleski}, {Steber}, {Pate}, {Lattanzi}, {Spezzano}, {McCarthy}, \& {Remijan}}]{Neill2012}
{Neill}, J.~L., {Muckle}, M.~T., {Zaleski}, D.~P., {et~al.} 2012, \apj, 755, 153, \dodoi{10.1088/0004-637X/755/2/153}

\bibitem[{Oró \& Kimball(1961)}]{Oro&Kimball1961}
Oró, J., \& Kimball, A. 1961, Archives of Biochemistry and Biophysics, 94, 217, \dodoi{https://doi.org/10.1016/0003-9861(61)90033-9}

\bibitem[{{Patel} {et~al.}(2015){Patel}, {Percivalle}, {Ritson}, {Duffy}, \& {Sutherland}}]{Patel2015}
{Patel}, B.~H., {Percivalle}, C., {Ritson}, D.~J., {Duffy}, C.~D., \& {Sutherland}, J.~D. 2015, Nature Chemistry, 7, 301, \dodoi{10.1038/nchem.2202}

\bibitem[{{Petrie} \& {Osamura}(2004)}]{Petrie2004}
{Petrie}, S., \& {Osamura}, Y. 2004, Journal of Physical Chemistry A, 108, 3623, \dodoi{10.1021/jp0378182}

\bibitem[{{Powner} {et~al.}(2009){Powner}, {Gerland}, \& {Sutherland}}]{Powner2009}
{Powner}, M.~W., {Gerland}, B., \& {Sutherland}, J.~D. 2009, \nat, 459, 239, \dodoi{10.1038/nature08013}

\bibitem[{{Puzzarini}(2015)}]{Puzzarini2015}
{Puzzarini}, C. 2015, Journal of Physical Chemistry A, 119, 11614, \dodoi{10.1021/acs.jpca.5b09489}

\bibitem[{{Puzzarini} \& {Barone}(2020)}]{Puzzarini&Barone2020}
{Puzzarini}, C., \& {Barone}, V. 2020, Physical Chemistry Chemical Physics (Incorporating Faraday Transactions), 22, 6507, \dodoi{10.1039/D0CP00561D}

\bibitem[{{Requena-Torres} {et~al.}(2008){Requena-Torres}, {Mart{\'\i}n-Pintado}, {Mart{\'\i}n}, \& {Morris}}]{Requena-Torres2008}
{Requena-Torres}, M.~A., {Mart{\'\i}n-Pintado}, J., {Mart{\'\i}n}, S., \& {Morris}, M.~R. 2008, \apj, 672, 352, \dodoi{10.1086/523627}

\bibitem[{{Requena-Torres} {et~al.}(2006){Requena-Torres}, {Mart{\'\i}n-Pintado}, {Rodr{\'\i}guez-Franco}, {Mart{\'\i}n}, {Rodr{\'\i}guez-Fern{\'a}ndez}, \& {de Vicente}}]{Requena-Torres2006}
{Requena-Torres}, M.~A., {Mart{\'\i}n-Pintado}, J., {Rodr{\'\i}guez-Franco}, A., {et~al.} 2006, \aap, 455, 971, \dodoi{10.1051/0004-6361:20065190}

\bibitem[{{Rimola} {et~al.}(2018){Rimola}, {Skouteris}, {Balucani}, {Ceccarelli}, {Enrique-Romero}, {Taquet}, \& {Ugliengo}}]{Rimola2018}
{Rimola}, A., {Skouteris}, D., {Balucani}, N., {et~al.} 2018, ACS Earth and Space Chemistry, 2, 720, \dodoi{10.1021/acsearthspacechem.7b00156}

\bibitem[{{Rivilla} {et~al.}(2017){Rivilla}, {Beltr{\'a}n}, {Mart{\'\i}n-Pintado}, {Fontani}, {Caselli}, \& {Cesaroni}}]{Rivilla2017}
{Rivilla}, V.~M., {Beltr{\'a}n}, M.~T., {Mart{\'\i}n-Pintado}, J., {et~al.} 2017, \aap, 599, A26, \dodoi{10.1051/0004-6361/201628823}

\bibitem[{{Rivilla} {et~al.}(2018){Rivilla}, {Jim{\'e}nez-Serra}, {Zeng}, {Mart{\'\i}n}, {Mart{\'\i}n-Pintado}, {Armijos-Abenda{\~n}o}, {Viti}, {Aladro}, {Riquelme}, {Requena-Torres}, {Qu{\'e}nard}, {Fontani}, \& {Beltr{\'a}n}}]{Rivilla2018}
{Rivilla}, V.~M., {Jim{\'e}nez-Serra}, I., {Zeng}, S., {et~al.} 2018, \mnras, 475, L30, \dodoi{10.1093/mnrasl/slx208}

\bibitem[{{Rivilla} {et~al.}(2019){Rivilla}, {Mart{\'i}n-Pintado}, {Jim{\'e}nez-Serra}, {Zeng}, {Mart{\'i}n}, {Armijos-Abenda{\~n}o}, {Requena-Torres}, {Aladro}, \& {Riquelme}}]{Rivilla2019}
{Rivilla}, V.~M., {Mart{\'i}n-Pintado}, J., {Jim{\'e}nez-Serra}, I., {et~al.} 2019, \mnras, 483, L114, \dodoi{10.1093/mnrasl/sly228}

\bibitem[{{Rivilla} {et~al.}(2020{\natexlab{a}}){Rivilla}, {Drozdovskaya}, {Altwegg}, {Caselli}, {Beltr{\'a}n}, {Fontani}, {van der Tak}, {Cesaroni}, {Vasyunin}, {Rubin}, {Lique}, {Marinakis}, {Testi}, {Rosina Team}, {Balsiger}, {Berthelier}, {de Keyser}, {Fiethe}, {Fuselier}, {Gasc}, {Gombosi}, {S{\'e}mon}, \& {Tzou}}]{Rivilla2020a}
{Rivilla}, V.~M., {Drozdovskaya}, M.~N., {Altwegg}, K., {et~al.} 2020{\natexlab{a}}, \mnras, 492, 1180, \dodoi{10.1093/mnras/stz3336}

\bibitem[{{Rivilla} {et~al.}(2020{\natexlab{b}}){Rivilla}, {Mart{\'\i}n-Pintado}, {Jim{\'e}nez-Serra}, {Mart{\'i}n}, {Rodr{\'i}guez-Almeida}, {Requena-Torres}, {Rico-Villas}, {Zeng}, \& {Briones}}]{Rivilla2020b}
{Rivilla}, V.~M., {Mart{\'\i}n-Pintado}, J., {Jim{\'e}nez-Serra}, I., {et~al.} 2020{\natexlab{b}}, \apjl, 899, L28, \dodoi{10.3847/2041-8213/abac55}

\bibitem[{{Rivilla} {et~al.}(2021{\natexlab{a}}){Rivilla}, {Jim{\'e}nez-Serra}, {Garc{\'i}a de la Concepci{\'o}n}, {Mart{\'i}n-Pintado}, {Colzi}, {Rodr{\'i}guez-Almeida}, {Tercero}, {Rico-Villas}, {Zeng}, {Mart{\'i}n}, {Requena-Torres}, \& {de Vicente}}]{Rivilla2021b}
{Rivilla}, V.~M., {Jim{\'e}nez-Serra}, I., {Garc{\'i}a de la Concepci{\'o}n}, J., {et~al.} 2021{\natexlab{a}}, \mnras, 506, L79, \dodoi{10.1093/mnrasl/slab074}

\bibitem[{{Rivilla} {et~al.}(2021{\natexlab{b}}){Rivilla}, {Jim{\'e}nez-Serra}, {Mart{\'i}n-Pintado}, {Briones}, {Rodr{\'i}guez-Almeida}, {Rico-Villas}, {Tercero}, {Zeng}, {Colzi}, {de Vicente}, {Mart{\'i}n}, \& {Requena-Torres}}]{Rivilla2021a}
{Rivilla}, V.~M., {Jim{\'e}nez-Serra}, I., {Mart{\'i}n-Pintado}, J., {et~al.} 2021{\natexlab{b}}, Proceedings of the National Academy of Science, 118, e2101314118, \dodoi{10.1073/pnas.2101314118}

\bibitem[{{Rivilla} {et~al.}(2022{\natexlab{a}}){Rivilla}, {Jim{\'e}nez-Serra}, {Mart{\'\i}n-Pintado}, {Colzi}, {Tercero}, {de Vicente}, {Zeng}, {Mart{\'\i}n}, {Garc{\'\i}a de la Concepci{\'o}n}, {Bizzocchi}, {Melosso}, {Rico-Villas}, \& {Requena-Torres}}]{Rivilla2022c}
{Rivilla}, V.~M., {Jim{\'e}nez-Serra}, I., {Mart{\'\i}n-Pintado}, J., {et~al.} 2022{\natexlab{a}}, Frontiers in Astronomy and Space Sciences, 9, 876870, \dodoi{10.3389/fspas.2022.876870}

\bibitem[{{Rivilla} {et~al.}(2022{\natexlab{b}}){Rivilla}, {Colzi}, {Jim{\'e}nez-Serra}, {Mart{\'i}n-Pintado}, {Meg{\'i}as}, {Melosso}, {Bizzocchi}, {L{\'o}pez-Gallifa}, {Mart{\'i}nez-Henares}, {Massalkhi}, {Tercero}, {de Vicente}, {Guillemin}, {Garc{\'i}a de la Concepci{\'o}n}, {Rico-Villas}, {Zeng}, {Mart{\'i}n}, {Requena-Torres}, {Tonolo}, {Alessandrini}, {Dore}, {Barone}, \& {Puzzarini}}]{Rivilla2022a}
{Rivilla}, V.~M., {Colzi}, L., {Jim{\'e}nez-Serra}, I., {et~al.} 2022{\natexlab{b}}, \apjl, 929, L11, \dodoi{10.3847/2041-8213/ac6186}

\bibitem[{{Rivilla} {et~al.}(2022{\natexlab{c}}){Rivilla}, {Garc{\'i}a De La Concepci{\'o}n}, {Jim{\'e}nez-Serra}, {Mart{\'i}n-Pintado}, {Colzi}, {Tercero}, {Meg{\'i}as}, {L{\'o}pez-Gallifa}, {Mart{\'i}nez-Henares}, {Massalkhi}, {Mart{\'i}n}, {Zeng}, {De Vicente}, {Rico-Villas}, {Requena-Torres}, \& {Cosentino}}]{Rivilla2022b}
{Rivilla}, V.~M., {Garc{\'i}a De La Concepci{\'o}n}, J., {Jim{\'e}nez-Serra}, I., {et~al.} 2022{\natexlab{c}}, Frontiers in Astronomy and Space Sciences, 9, 829288, \dodoi{10.3389/fspas.2022.829288}

\bibitem[{{Rivilla} {et~al.}(2023){Rivilla}, {Sanz-Novo}, {Jim{\'e}nez-Serra}, {Mart{\'\i}n-Pintado}, {Colzi}, {Zeng}, {Meg{\'\i}as}, {L{\'o}pez-Gallifa}, {Mart{\'\i}nez-Henares}, {Massalkhi}, {Tercero}, {de Vicente}, {Mart{\'\i}n}, {San Andr{\'e}s}, {Requena-Torres}, \& {Alonso}}]{Rivilla2023}
{Rivilla}, V.~M., {Sanz-Novo}, M., {Jim{\'e}nez-Serra}, I., {et~al.} 2023, \apjl, 953, L20, \dodoi{10.3847/2041-8213/ace977}

\bibitem[{{Rodr{\'i}guez-Almeida} {et~al.}(2021{\natexlab{a}}){Rodr{\'i}guez-Almeida}, {Jim{\'e}nez-Serra}, {Rivilla}, {Mart{\'i}n-Pintado}, {Zeng}, {Tercero}, {de Vicente}, {Colzi}, {Rico-Villas}, {Mart{\'i}n}, \& {Requena-Torres}}]{Rodriguez-Almeida2021a}
{Rodr{\'i}guez-Almeida}, L.~F., {Jim{\'e}nez-Serra}, I., {Rivilla}, V.~M., {et~al.} 2021{\natexlab{a}}, \apjl, 912, L11, \dodoi{10.3847/2041-8213/abf7cb}

\bibitem[{{Rodr{\'i}guez-Almeida} {et~al.}(2021{\natexlab{b}}){Rodr{\'i}guez-Almeida}, {Rivilla}, {Jim{\'e}nez-Serra}, {Melosso}, {Colzi}, {Zeng}, {Tercero}, {de Vicente}, {Mart{\'i}n}, {Requena-Torres}, {Rico-Villas}, \& {Mart{\'i}n-Pintado}}]{Rodriguez-Almeida2021b}
{Rodr{\'i}guez-Almeida}, L.~F., {Rivilla}, V.~M., {Jim{\'e}nez-Serra}, I., {et~al.} 2021{\natexlab{b}}, \aap, 654, L1, \dodoi{10.1051/0004-6361/202141989}

\bibitem[{{San Andr{\'e}s} {et~al.}(2023){San Andr{\'e}s}, {Colzi}, {Rivilla}, {Garc{\'i}a de la Concepci{\'o}n}, {Melosso}, {Mart{\'i}n-Pintado}, {Jim{\'e}nez-Serra}, {Zeng}, {Mart{\'i}n}, \& {Requena-Torres}}]{SanAndres2023}
{San Andr{\'e}s}, D., {Colzi}, L., {Rivilla}, V.~M., {et~al.} 2023, \mnras, 523, 3239, \dodoi{10.1093/mnras/stad1385}

\bibitem[{Sanchez {et~al.}(1968)Sanchez, Ferris, \& Orgel}]{Sanchez1968}
Sanchez, R.~A., Ferris, J.~P., \& Orgel, L.~E. 1968, Journal of Molecular Biology, 38, 121, \dodoi{https://doi.org/10.1016/0022-2836(68)90132-0}

\bibitem[{{Sandstr{\"o}m} \& {Rahm}(2023)}]{Sandstrom&Rahm2023}
{Sandstr{\"o}m}, H., \& {Rahm}, M. 2023, Journal of Physical Chemistry A, 127, 4503, \dodoi{10.1021/acs.jpca.3c01504}

\bibitem[{{Santalucia} {et~al.}(2022){Santalucia}, {Pazzi}, {Bonino}, {Signorile}, {Scarano}, {Ugliengo}, {Spoto}, \& {Mino}}]{Santalucia2022}
{Santalucia}, R., {Pazzi}, M., {Bonino}, F., {et~al.} 2022, Physical Chemistry Chemical Physics (Incorporating Faraday Transactions), 24, 7224, \dodoi{10.1039/D1CP05407D}

\bibitem[{{Sanz-Novo} {et~al.}(2023){Sanz-Novo}, {Rivilla}, {Jim{\'e}nez-Serra}, {Mart{\'\i}n-Pintado}, {Colzi}, {Zeng}, {Meg{\'\i}as}, {L{\'o}pez-Gallifa}, {Mart{\'\i}nez-Henares}, {Massalkhi}, {Tercero}, {de Vicente}, {Mart{\'\i}n}, {San Andr{\'e}s}, \& {Requena-Torres}}]{Sanz-Novo2023}
{Sanz-Novo}, M., {Rivilla}, V.~M., {Jim{\'e}nez-Serra}, I., {et~al.} 2023, \apj, 954, 3, \dodoi{10.3847/1538-4357/ace523}

\bibitem[{{Shingledecker} {et~al.}(2019){Shingledecker}, {{\'A}lvarez-Barcia}, {Korn}, \& {K{\"a}stner}}]{Shingledecker2019}
{Shingledecker}, C.~N., {{\'A}lvarez-Barcia}, S., {Korn}, V.~H., \& {K{\"a}stner}, J. 2019, \apj, 878, 80, \dodoi{10.3847/1538-4357/ab1d4a}

\bibitem[{{Shingledecker} {et~al.}(2020){Shingledecker}, {Molpeceres}, {Rivilla}, {Majumdar}, \& {K{\"a}stner}}]{Shingledecker2020}
{Shingledecker}, C.~N., {Molpeceres}, G., {Rivilla}, V.~M., {Majumdar}, L., \& {K{\"a}stner}, J. 2020, \apj, 897, 158, \dodoi{10.3847/1538-4357/ab94b5}

\bibitem[{{Shivani} {et~al.}(2017){Shivani}, {Misra}, \& {Tandon}}]{Shivani2017}
{Shivani}, {Misra}, A., \& {Tandon}, P. 2017, Research in Astronomy and Astrophysics, 17, 1, \dodoi{10.1088/1674-4527/17/1/1}

\bibitem[{Skodje {et~al.}(1981)Skodje, Truhlar, \& Garrett}]{Skodje1981}
Skodje, R.~T., Truhlar, D.~G., \& Garrett, B.~C. 1981, The Journal of Physical Chemistry, 85, 3019, \dodoi{10.1021/j150621a001}

\bibitem[{{Smith} {et~al.}(2001){Smith}, {Talbi}, \& {Herbst}}]{Smith2001}
{Smith}, I.~W.~M., {Talbi}, D., \& {Herbst}, E. 2001, \aap, 369, 611, \dodoi{10.1051/0004-6361:20010126}

\bibitem[{{Stolze} {et~al.}(1989){Stolze}, {Sutter}, \& {Wentrup}}]{Stolze1989}
{Stolze}, W.~H., {Sutter}, D.~H., \& {Wentrup}, C. 1989, Zeitschrift Naturforschung Teil A, 44, 291, \dodoi{10.1515/zna-1989-0407}

\bibitem[{{Suzuki} {et~al.}(2016){Suzuki}, {Ohishi}, {Hirota}, {Saito}, {Majumdar}, \& {Wakelam}}]{Suzuki2016}
{Suzuki}, T., {Ohishi}, M., {Hirota}, T., {et~al.} 2016, \apj, 825, 79, \dodoi{10.3847/0004-637X/825/1/79}

\bibitem[{{Takano} {et~al.}(1990){Takano}, {Sugie}, {Sugawara}, {Takeo}, {Matsumura}, {Masuda}, \& {Kuchitsu}}]{Takano1990}
{Takano}, S., {Sugie}, M., {Sugawara}, K.-i., {et~al.} 1990, Journal of Molecular Spectroscopy, 141, 13, \dodoi{10.1016/0022-2852(90)90273-S}

\bibitem[{{Tercero} {et~al.}(2021){Tercero}, {L{\'o}pez-P{\'e}rez}, {Gallego}, {Beltr{\'a}n}, {Garc{\'i}a}, {Patino-Esteban}, {L{\'o}pez-Fern{\'a}ndez}, {G{\'o}mez-Molina}, {Diez}, {Garc{\'i}a-Carre{\~n}o}, {Malo}, {Amils}, {Serna}, {Albo}, {Hern{\'a}ndez}, {Vaquero}, {Gonz{\'a}lez-Garc{\'i}a}, {Barbas}, {L{\'o}pez-Fern{\'a}ndez}, {Bujarrabal}, {G{\'o}mez-Garrido}, {Pardo}, {Santander-Garc{\'i}a}, {Tercero}, {Cernicharo}, \& {de Vicente}}]{Tercero2021}
{Tercero}, F., {L{\'o}pez-P{\'e}rez}, J.~A., {Gallego}, J.~D., {et~al.} 2021, \aap, 645, A37, \dodoi{10.1051/0004-6361/202038701}

\bibitem[{{Theule} {et~al.}(2011){Theule}, {Borget}, {Mispelaer}, {Danger}, {Duvernay}, {Guillemin}, \& {Chiavassa}}]{Theule2011}
{Theule}, P., {Borget}, F., {Mispelaer}, F., {et~al.} 2011, \aap, 534, A64, \dodoi{10.1051/0004-6361/201117494}

\bibitem[{{Turner} {et~al.}(1975){Turner}, {Liszt}, {Kaifu}, \& {Kisliakov}}]{Turner1975}
{Turner}, B.~E., {Liszt}, H.~S., {Kaifu}, N., \& {Kisliakov}, A.~G. 1975, \apjl, 201, L149, \dodoi{10.1086/181963}

\bibitem[{{Vasconcelos} {et~al.}(2020){Vasconcelos}, {Pilling}, {Agnihotri}, {Rothard}, \& {Boduch}}]{Vasconcelos2020}
{Vasconcelos}, F.~A., {Pilling}, S., {Agnihotri}, A., {Rothard}, H., \& {Boduch}, P. 2020, \icarus, 351, 113944, \dodoi{10.1016/j.icarus.2020.113944}

\bibitem[{{Vazart} {et~al.}(2015){Vazart}, {Latouche}, {Skouteris}, {Balucani}, \& {Barone}}]{Vazart2015}
{Vazart}, F., {Latouche}, C., {Skouteris}, D., {Balucani}, N., \& {Barone}, V. 2015, \apj, 810, 111, \dodoi{10.1088/0004-637X/810/2/111}

\bibitem[{{Winnewisser} {et~al.}(1984){Winnewisser}, {Winnewisser}, \& {Wentrup}}]{Winnewisser1984}
{Winnewisser}, M., {Winnewisser}, B.~P., \& {Wentrup}, C. 1984, Journal of Molecular Spectroscopy, 105, 193, \dodoi{10.1016/0022-2852(84)90111-5}

\bibitem[{{Yim} \& {Choe}(2012)}]{Yim&Choe2012}
{Yim}, M.~K., \& {Choe}, J.~C. 2012, Chemical Physics Letters, 538, 24, \dodoi{10.1016/j.cplett.2012.04.042}

\bibitem[{{Zaleski} {et~al.}(2013){Zaleski}, {Seifert}, {Steber}, {Muckle}, {Loomis}, {Corby}, {Martinez}, {Crabtree}, {Jewell}, {Hollis}, {Lovas}, {Vasquez}, {Nyiramahirwe}, {Sciortino}, {Johnson}, {McCarthy}, {Remijan}, \& {Pate}}]{Zaleski2013}
{Zaleski}, D.~P., {Seifert}, N.~A., {Steber}, A.~L., {et~al.} 2013, \apjl, 765, L10, \dodoi{10.1088/2041-8205/765/1/L10}

\bibitem[{{Zeng} {et~al.}(2019){Zeng}, {Qu{\'e}nard}, {Jim{\'e}nez-Serra}, {Mart{\'i}n-Pintado}, {Rivilla}, {Testi}, \& {Mart{\'i}n-Dom{\'e}nech}}]{Zeng2019}
{Zeng}, S., {Qu{\'e}nard}, D., {Jim{\'e}nez-Serra}, I., {et~al.} 2019, \mnras, 484, L43, \dodoi{10.1093/mnrasl/slz002}

\bibitem[{{Zeng} {et~al.}(2018){Zeng}, {Jim{\'e}nez-Serra}, {Rivilla}, {Mart{\'i}n}, {Mart{\'i}n-Pintado}, {Requena-Torres}, {Armijos-Abenda{\~n}o}, {Riquelme}, \& {Aladro}}]{Zeng2018}
{Zeng}, S., {Jim{\'e}nez-Serra}, I., {Rivilla}, V.~M., {et~al.} 2018, \mnras, 478, 2962, \dodoi{10.1093/mnras/sty1174}

\bibitem[{{Zeng} {et~al.}(2020){Zeng}, {Zhang}, {Jim{\'e}nez-Serra}, {Tercero}, {Lu}, {Mart{\'i}n-Pintado}, {de Vicente}, {Rivilla}, \& {Li}}]{Zeng2020}
{Zeng}, S., {Zhang}, Q., {Jim{\'e}nez-Serra}, I., {et~al.} 2020, \mnras, 497, 4896, \dodoi{10.1093/mnras/staa2187}

\bibitem[{{Zeng} {et~al.}(2021){Zeng}, {Jim{\'e}nez-Serra}, {Rivilla}, {Mart{\'i}n-Pintado}, {Rodr{\'i}guez-Almeida}, {Tercero}, {de Vicente}, {Rico-Villas}, {Colzi}, {Mart{\'i}n}, \& {Requena-Torres}}]{Zeng2021}
{Zeng}, S., {Jim{\'e}nez-Serra}, I., {Rivilla}, V.~M., {et~al.} 2021, \apjl, 920, L27, \dodoi{10.3847/2041-8213/ac2c7e}

\bibitem[{{Zeng} {et~al.}(2023){Zeng}, {Rivilla}, {Jim{\'e}nez-Serra}, {Colzi}, {Mart{\'i}n-Pintado}, {Tercero}, {de Vicente}, {Mart{\'i}n}, \& {Requena-Torres}}]{Zeng2023}
{Zeng}, S., {Rivilla}, V.~M., {Jim{\'e}nez-Serra}, I., {et~al.} 2023, \mnras, 523, 1448, \dodoi{10.1093/mnras/stad1478}

\bibitem[{{Zhang} {et~al.}(2020){Zhang}, {Quan}, {Chang}, {Herbst}, {Esimbek}, \& {Webb}}]{Zhang2020}
{Zhang}, X., {Quan}, D., {Chang}, Q., {et~al.} 2020, \mnras, 497, 609, \dodoi{10.1093/mnras/staa1979}

\bibitem[{{Zheng} {et~al.}(2024){Zheng}, {Li}, {Wang}, {Wang}, {Gao}, {Quan}, {Du}, {Wu}, {Bergin}, \& {Li}}]{Zheng2024}
{Zheng}, S., {Li}, J., {Wang}, J., {et~al.} 2024, \apj, 961, 58, \dodoi{10.3847/1538-4357/ad072c}

\end{thebibliography}
\bibliographystyle{aasjournal}

%% For this sample we use BibTeX plus aasjournals.bst to generate the
%% the bibliography. The sample63.bib file was populated from ADS. To
%% get the citations to show in the compiled file do the following:
%%
%% pdflatex sample63.tex
%% bibtext sample63
%% pdflatex sample63.tex
%% pdflatex sample63.tex

%\newpage
\appendix

\section{\ce{H2CNCN} transitions excluded from the LTE fit}
\label{appendix:H2CNCN_blended_transitions}

\begin{figure*}[h]
    \centering
    \includegraphics[width=\textwidth]{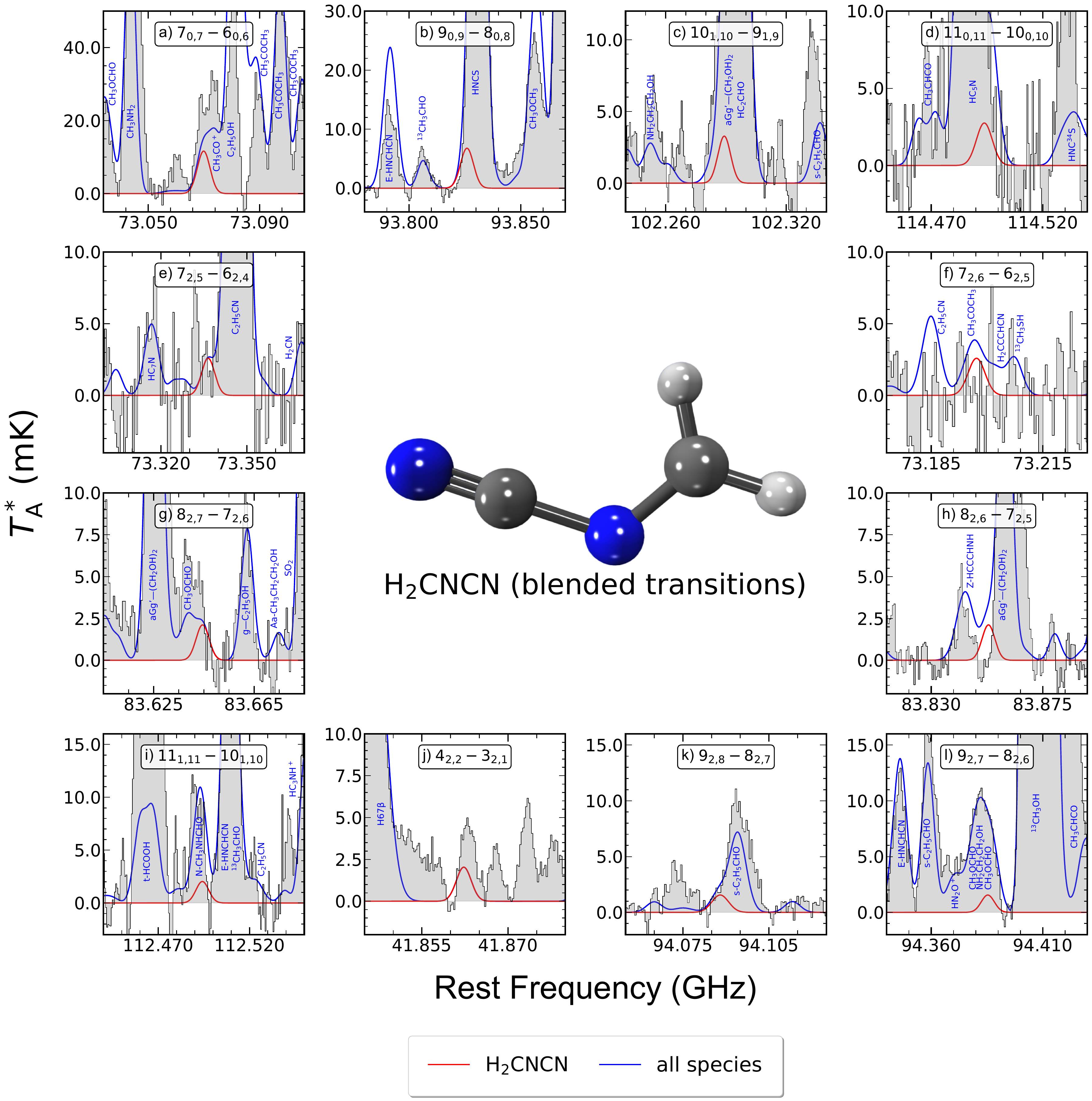}
    \caption{\ce{H2CNCN} rotational transitions targeted towards G+0.693 but excluded when performing the LTE fit of this molecule, which predicts the emission profiles delineated by the red solid lines. Black histogram and grey-shaded areas indicate the observed spectrum, while the blue solid lines represent the emission of all the species already identified in the cloud (whose names are indicated by the blue labels). Transitions shown here are sorted by decreasing peak intensity, which goes down to 1\mK. Panel labels indicate the main rotational transition being targeted using the $J_{K_\text{a}K_\text{c}}$ common notation.} 
    \label{fig:H2CNCN_blended_transitions} 
\end{figure*}

Besides \ce{H2CNCN} transitions used to generate the LTE model of this molecule shown in Fig.~\ref{fig:H2CNCN_detected_lines}, our observational data also encompasses many other $a$-type rotational lines belonging to this species within the almost entirely targeted $J_\text{up}$ = 3 to $J_\text{up} = 11$ progressions of the $K_a = 0, 1, 2$ ladders. These remaining transitions are shown in Fig.~\ref{fig:H2CNCN_blended_transitions}, which just includes the most intense ones among them (with peak intensities down to 1.5\mK and integrated S/N$>$3). As it can be seen, most of these lines are heavily blended with the emission from other species (either unknown or already detected towards G+0.693), while a few others lie in certain segments of the spectrum where the noise level is too high to provide a clear identification. Since it is evident that these transitions do not qualify for the direct detection of this molecule, they were excluded in our analysis. Nevertheless, it is worth emphasising that the predicted emissions for all of them is coherent with the observed spectrum, reaffirming the consistency of the LTE model derived for \ce{H2CNCN} as presented in Sect.~\ref{subsec:detection_of_H2CNCN}.

\section{$Z$-\ce{HNCHCN} and $E$-\ce{HNCHCN} spectroscopy} \label{appendix:Z,E-HNCHCN_spectroscopy}

Table~\ref{tab:Z,E-HNCHCN_spectroscopy} gathers the spectroscopic information of the lines belonging to $Z$-\ce{HNCHCN} and $E$-\ce{HNCHCN} we have detected towards G+0.693 that are unblended or partially blended with the emission of other molecular species. These transitions are shown in Figs.~\ref{fig:Z-HNCHCN_detection} and \ref{fig:E-HNCHCN_detection}, respectively, and are those we used to perform the LTE line fitting of these two isomers. We provide the most relevant spectroscopic parameters for each transition, which are the rest frequency, quantum numbers, base 10 logarithm of the integrated intensity at 300\K ($\log I$), energy of the upper level ($E_\text{up}$), the noise level of the spectra evaluated over line-free spectral ranges close to it (rms), the integrated intensity ($\int{T^*_\text{A}\text{d}v}$) as calculated from the fit, the detection level (in terms of the integrated S/N ratio) and information regarding line blending. The spectroscopic information has been obtained from the CDMS entries of these isomers, which come from \citet{Takano1990, Zaleski2013} and \citet{Melosso2018} works. 

\clearpage
\onecolumngrid
\startlongtable
\begin{deluxetable}{DcDDcccc}
\tabletypesize{\footnotesize}
\tablenum{3}
\setlength{\tabcolsep}{2.9pt}
\tablecaption{Spectroscopic information of $Z$- and $E$-\ce{HNCHCN} unblended and partially blended transitions detected in G+0.693.\label{tab:Z,E-HNCHCN_spectroscopy}}
\tablehead{
\multicolumn2c{Frequency$^{(a)}$} & \colhead{Transition$^{(b)}$} & \multicolumn2c{$\log I$} & \multicolumn2c{$E_\text{up}$} & \colhead{rms} & \colhead{$\int{T^*_\text{A}\text{d}v}$} & \colhead{S/N$^{(c)}$} & \colhead{Blending$^{(d)}$} \\
\multicolumn2c{($\text{GHz}$)} & \colhead{($J_{K_\text{a},K_\text{c}}$$'-$$J_{K_\text{a},K_\text{c}}$$''$)} & \multicolumn2c{($\text{nm}^2 \, \text{MHz}$)} & \multicolumn2c{($\text{K}$)} & \colhead{($\text{mK}$)} & \colhead{($\text{mK}\,\text{km}\,\text{s}^{-1}$)} & \colhead{} & \colhead{}
}
\decimals
\startdata
\multicolumn{11}{c}{$Z$-\ce{HNCHCN}} \\
\hline
37.93654(7) & $4_{1,4}-3_{1,3}$ & -6.3795 & 6.9 & 0.5 & 236(14) & 78(5) & Blended: \ce{C2H3CN} \\
38.7947810(4) & $4_{0,4}-3_{0,3}$ & -6.3287 & 4.7 & 0.5 & 302(15) & 100(5) & Slightly blended: \ce{CH3COCH3} \\
39.70212(7) & $4_{1,3}-3_{1,2}$ & -6.3402 & 7.1 & 0.5 & 244(14) & 82(5) & Blended: \ce{CH3CONH2} and U \\
47.4148090(8) & $5_{1,5}-4_{1,4}$ & -6.1065 & 9.2 & 0.5 & 325(16) & 119(6) & Blended: \ce{C2H3CN} \\
48.47090(7) & $5_{0,5}-4_{0,4}$ & -6.0665 & 7.0 & 0.5 & 405(18) & 151(7) & Unblended \\
48.5276164(5) & $5_{2,4}-4_{2,3}$ & -6.1548 & 16.5 & 0.5 & 174(12) & 65(5) & Slightly blended: \ce{HOCH2CHO} and U \\
49.62143(7) & $5_{1,4}-4_{1,3}$ & -6.0674 & 9.5 & 0.5 & 334(16) & 125(6) & Unblended \\
75.8241226(12) & $8_{1,8}-7_{1,7}$ & -5.5395 & 18.7 & 2.4 & 447(60) & 26(3) & Blended: \ce{CH3SH} and U \\
77.3975188(8) & $8_{0,8}-7_{0,7}$ & -5.5120 & 16.7 & 2.4 & 536(60) & 31(3) & Unblended$^\star$ \\
77.61551(7) & $8_{2,7}-7_{2,6}$ & -5.5512 & 26.2 & 2.4 & 257(60) & 15(3) & Blended: $cis$-$N$-\ce{CH3NHCHO} \\
77.86298(7) & $8_{2,6}-7_{2,5}$ & -5.5485 & 26.3 & 2.4 & 257(60) & 15(3) & Unblended$^\star$ \\
85.2831815(13) & $9_{1,9}-8_{1,8}$$^*$ & -5.3992 & 22.8 & 1.6 & 429(40) & 39(4) & Slightly blended: \ce{HOCH2CHO} and \ce{H^{15}NCO} \\
86.99658(10) & $9_{0,9}-8_{0,8}$$^*$ & -5.3738 & 20.9 & 1.6 & 508(41) & 47(4) & Blended: \ce{HCCCC^{13}CN} \\
87.6558177(9) & $9_{2,7}-8_{2,6}$ & -5.4029 & 30.5 & 1.3 & 247(31) & 28(4) & Blended: $aGg'$-\ce{(CH2OH)2} \\
89.24785(10) & $9_{1,8}-8_{1,7}$$^*$ & -5.3610 & 23.8 & 1.3 & 421(34) & 49(4) & Blended: \ce{H2CCCHCN} and U \\
94.73583(10) & $10_{1,10}-9_{1,9}$$^*$ & -5.2746 & 27.4 & 0.9 & 388(25) & 66(4) & Unblended \\
96.56986(10) & $10_{0,10}-9_{0,9}$$^*$ & -5.2510 & 25.5 & 1.5 & 455(40) & 47(4) & Slightly blended: \ce{^{13}CH3SH} \\
97.46902(10) & $10_{2,8}-9_{2,7}$ & -5.2744 & 35.1 & 1.5 & 223(35) & 23(4) & Blended: \ce{CH3CONH2} and U \\
99.13721(10) & $10_{1,9}-9_{1,8}$ & -5.2367 & 28.6 & 1.5 & 375(40) & 39(4) & Blended: \ce{CH3NH2} and \ce{CH3OCHO} \\
104.181786(20) & $11_{1,11}-10_{1,10}$ & -5.1629 & 32.4 & 1.8 & 332(43) & 30(4) & Slightly blended: $aGg'$-\ce{(CH2OH)2} \\
106.115423(20) & $11_{0,11}-10_{0,10}$ & -5.1409 & 30.6 & 1.8 & 387(44) & 35(4) & Blended: \ce{CH3OCHO} \\
106.663856(20) & $11_{2,10}-10_{2,9}$ & -5.1647 & 40.2 & 1.8 & 191(42) & 17(4) & Blended: $aGg'$-\ce{(CH2OH)2} \\
109.016976(20) & $11_{1,10}-10_{1,9}$$^*$ & -5.1254 & 33.8 & 2.9 & 316(70) & 18(4) & Slightly blended with U \\
113.620315(20) & $12_{1,12}-11_{1,11}$ & -5.0620 & 37.8 & 2.8 & 271(70) & 16(4) & Blended: $Aa$-\ce{CH3CH2CH2OH} and $g$-\ce{C2H5SH} \\
125.1163734(14) & $13_{0,13}-12_{0,12}$ & -4.9511 & 42.2 & 6.6 & 241(153) & 7(4) & Blended: \ce{CH3SH} \\
134.5699312(15) & $14_{0,14}-13_{0,13}$ & -4.8686 & 48.7 & 1.5 & 177(35) & 22(4) & Unblended$^\star$ \\
\hline
\multicolumn{11}{c}{$E$-\ce{HNCHCN}} \\
\hline
37.54275(10) & $4_{1,4}-3_{1,3}$ & -5.6004 & 7.3 & 0.5 & 297(13) & 97(5) & Slightly blended with U \\
46.9248305(11) & $5_{1,5}-4_{1,4}$ & -5.3274 & 9.5 & 0.5 & 415(16) & 152(6) & Unblended$^\star$ \\
48.78317(10) & $5_{1,4}-4_{1,3}$ & -5.2940 & 9.8 & 0.5 & 425(16) & 159(6) & Unblended$^\star$ \\
75.05519(10) & $8_{1,8}-7_{1,7}$ & -4.7600 & 19.0 & 2.4 & 606(60) & 35(3) & Unblended$^\star$ \\
78.02648(10) & $8_{1,7}-7_{1,6}$ & -4.7272 & 19.6 & 2.4 & 606(60) & 36(3) & Blended: \ce{CH3NCO} and U \\
84.42511(10) & $9_{1,9}-8_{1,8}$$^*$ & -4.6196 & 23.0 & 1.6 & 596(41) & 54(4) & Unblended \\
85.512670(16) & $3_{1,3}-2_{0,2}$ & -5.3412 & 5.5 & 1.6 & 256(40) & 23(3) & Blended: $aGg'$-\ce{(CH2OH)2} and U \\
87.76650(10) & $9_{1,8}-8_{1,7}$ & -4.5869 & 23.8 & 1.3 & 589(34) & 67(4) &  Blended: \ce{HOCH2CHO} and \ce{CH3OCHO} \\
93.79107(10) & $10_{1,10}-9_{1,9}$$^*$ & -4.4949 & 27.5 & 0.9 & 554(26) & 94(5) & Unblended \\
94.345774(16) & $4_{1,4}-3_{0,3}$ & -5.1964 & 7.3 & 0.9 & 317(23) & 54(4) & Unblended$^\star$ \\
95.42252(10) & $10_{0,10}-9_{0,9}$$^*$ & -4.4722 & 25.2 & 1.5 & 663(40) & 68(4) & Blended with U\\
97.50186(10) & $10_{1,9}-9_{1,8}$$^*$ & -4.4625 & 28.5 & 1.5 & 542(40) & 57(4) & Unblended \\
102.999619(16) & $5_{1,5}-4_{0,4}$ & -5.0669 & 9.5 & 1.8 & 362(43) & 32(4) & Blended: $l$-\ce{C3H2} and \ce{CH3COCH3} \\
103.1527434(22) & $11_{1,11}-10_{1,10}$ & -4.3830 & 32.5 & 1.8 & 489(44) & 44(4) & Unblended \\
104.895725(20) & $11_{0,11}-10_{0,10}$$^*$ & -4.3617 & 30.3 & 1.8 & 581(45) & 52(4) & Unblended$^\star$ \\
107.231510(20) & $11_{1,10}-10_{1,9}$ & -4.3509 & 33.7 & 1.8 & 473(44) & 43(4) & Unblended$^\star$ \\
111.479187(16) & $6_{1,6}-5_{0,5}$ & -4.9503 & 12.2 & 2.8 & 389(70) & 23(4) & Unblended$^\star$ \\
112.509636(20) & $12_{1,12}-11_{1,11}$ & -4.2819 & 37.9 & 2.8 & 413(70) & 25(4) & Blended: $E$-\ce{CH3CHNH} \\
114.349931(20) & $12_{0,12}-11_{0,11}$ & -4.2617 & 35.7 & 2.8 & 486(70) & 29(4) & Unblended \\
127.942187(15) & $8_{1,8}-7_{0,7}$ & -3.8856 & 19.0 & 6.6 & 383(160) & 10(4) & Blended: \ce{H^{15}NCO} and $Z$-\ce{CH3CHNH} \\
135.942373(15) & $9_{1,9}-8_{0,8}$ & -3.7881 & 23.0 & 1.5 & 356(40) & 44(5) & Blended: \ce{CH3OCHO} \\
\enddata
\tablecomments{ $^{(a)}$ The numbers in brackets represent the experimental uncertainty associated to the last digits, as measured by \citet{Takano1990}, \citet{Zaleski2013} and \citet{Melosso2018}. \\
$^{(b)}$ Rotational transitions marked with an asterisk symbol (*) refer to those previously targeted by \citet{Rivilla2019} and which these authors used to perform the LTE modelling of these species. All of the transitions are also labelled following the same notation indicated in Table~\ref{tab:H2CNCN_spectroscopy}. \\
$^{(c)}$ The S/N is calculated from the integrated intensity over the line width ($\int{T^*_\text{A}\text{d}v}$) and noise level $\sigma = \text{rms}\times\sqrt{\updelta v\times\text{FWHM}}$, where $\updelta v$ is the spectral resolution of the spectra in velocity units and the $\text{FWHM}$ is estimated from the LTE line fitting. The numbers in brackets represent the combined standard uncertainty associated to the last digits. \\
$^{(d)}$ The term ``unblended'' denotes those transitions which exhibit no contamination from other molecular species, while a star symbol ($\star$) is added when scarce line blending accounting for less than 5\% of the line total integrated intensity is present. ``U'' hints at blending with an unknown (not yet identified) species.}
\end{deluxetable}

\onecolumngrid

%% This command is needed to show the entire author+affiliation list when
%% the collaboration and author truncation commands are used.  It has to
%% go at the end of the manuscript.
%\allauthors

%% Include this line if you are using the \added, \replaced, \deleted
%% commands to see a summary list of all changes at the end of the article.
%\listofchanges

\end{document}